\documentclass[a4paper,fleqn,usenatbib]{mnras}

\usepackage{amsmath}
\usepackage{txfonts}

\usepackage[T1]{fontenc}
\usepackage{ae,aecompl}

\usepackage{graphicx}	
\usepackage{amsmath}	
\usepackage{amssymb}	

\title[Nature and statistical properties of quasar NALs]{Nature and statistical properties of quasar associated absorption systems in the XQ-100 Legacy Survey}

\author[S. Perrotta, V. D'Odorico, J. X. Prochaska et al.]{
S. Perrotta,$^{1,2,3}$\thanks{E-mail: sperrotta@sissa.it (SP)}
V. D'Odorico,$^{2}$
J. X. Prochaska,$^{3}$
S. Cristiani,$^{2,4}$
G. Cupani,$^{2}$
\newauthor
S. Ellison,$^{5}$
S. L\'{o}pez,$^{6}$
G. D. Becker,$^{7,8}$
T. A. M. Berg$^{5}$
L. Christensen,$^{9}$
K. D. Denney,$^{10}$
\newauthor
F. Hamann,$^{11,12}$
I. P\^{a}ris,$^{2}$
M. Vestergaard $^{9,13}$
and G. Worseck$^{14}$
\\
\\
$^{1}$SISSA - International School for Advanced Studies, Via Bonomea 265, 34136, Trieste, Italy\\
$^{2}$INAF-OATS, Via Tiepolo 11, 34143 Trieste, Italy\\
$^{3}$Department of Astronomy and Astrophysics, UCO/Lick Observatory, University of California, 1156 High Street, 16 Santa Cruz,\\ CA 95064, USA\\
$^{4}$ INFN / National Institute for Nuclear Physics, Via Valerio 2, I-34127 Trieste, Italy\\
$^{5}$Department of Physics and Astronomy, University of Victoria, Victoria, BC V8P 1A1, Canada\\
$^{6}$Departamento de Astronom\'ia, Universidad de Chile, Casilla 36-D, Santiago, Chile\\
$^{7}$Space Telescope Science Institute, 3700 San Martin Drive, Baltimore, MD 21218, USA\\
$^{8}$Institute of Astronomy and Kavli Institute of Cosmology, Madingley Road, Cambridge CB3 0HA, UK\\
$^{9}$Dark Cosmology Centre, Niels Bohr Institute, University of Copenhagen, Juliane Maries Vej 30, DK-2100 Copenhagen, Denmark\\
$^{10}$Department of Astronomy, The Ohio State University, 140 West 18th Avenue, Columbus, OH 43210, USA\\
$^{11}$Department of Astronomy, University of Florida, Gainesville, FL 32611-2055, USA\\
$^{12}$Department of Physics and Astronomy, University of California, 900 University Avenue, Riverside, CA 92521, USA\\
$^{13}$Department of Astronomy and Steward Observatory University of Arizona, 933 N Cherry Avenue Tucson AZ 85721,
USA\\
$^{14}$Max-Planck-Institut f$\rm \ddot{u}$r Astronomie, K$\rm \ddot{o}$nigstuhl 17, D-69117 Heidelberg, Germany\\
}

\date{Accepted XXX. Received YYY; in original form ZZZ}

\pubyear{2016}

\begin{document}
\label{firstpage}
\pagerange{\pageref{firstpage}--\pageref{lastpage}}
\maketitle

\begin{abstract}
We statistically study the physical properties of a sample of narrow absorption line (NAL) systems looking for empirical evidences to distinguish between intrinsic and intervening NALs without taking into account any a priori definition or velocity cut-off. 
We analyze the spectra of 100 quasars with $\rm 3.5 < z_{em} < 4.5$, observed with X-shooter/VLT in the context of the XQ-100 Legacy Survey.
We detect a $\rm \sim 8\,\sigma$ excess in the CIV number density within $\rm 10,000\;km\,s^{-1}$ of the quasar emission redshift with respect to the random occurrence of NALs. This excess does not show a dependence on the quasar bolometric luminosity and it is not due to the redshift evolution of NALs. It extends far beyond the standard 5000 $\rm km\,s^{-1}$ cut-off traditionally defined for {\it associated} absorption lines. We propose to modify this definition, extending the threshold to 10,000 $\rm km\,s^{-1}$  when also weak absorbers  ($\rm equivalent \,width<0.2$ \AA) \,are considered.
We infer NV is the ion that better traces the effects of the quasar ionization field, offering the best statistical tool to identify intrinsic systems. Following this criterion we estimate that the fraction of quasars in our sample hosting an intrinsic NAL system is 33 percent.
Lastly, we compare the properties of the material along the quasar line of sight, derived from our sample, with results based on close quasar pairs investigating the transverse direction. We find a deficiency of cool gas (traced by CII) along the line of sight connected to the quasar host galaxy, in contrast with what is observed in the transverse direction. 
\end{abstract}




\section{Introduction}

Supermassive black holes (SMBHs) are ubiquitous at the center of stellar spheroids. Moreover, their mass is tightly related to global properties of the host galaxy (e.g., Kormendy \& Ho 2013; McConnell \& Ma 2013; Ferrarese \& Merritt 2000). Despite this close interplay between SMBHs and their host systems, several compelling questions remain unanswered. Hence, a full description of why, how, and when black holes alter the evolutionary pathways of their host systems remain key questions in galaxy formation. High velocity quasar outflows appear to be a natural byproduct of accretion onto the SMBH and have therefore attracted much attention as a mechanism that can physically couple quasars to the evolution of their host galaxies.

Simulations of galaxy evolution (e.g., Booth \& Schaye 2009; Sijacki et al. 2015; Schaye et al. 2015) suggest that feedback from active galactic nuclei (AGN) plays a crucial role in heating the interstellar medium (ISM), quenching star formation and preventing massive galaxies to over-grow. As a consequence, the colors of the host galaxies evolve quickly and become red, in agreement with the observed color distribution of nearby galaxies.  
Quasar outflows may also contribute to the blowout of gas and dust from young galaxies, and thereby provide a mechanism for enriching the intergalactic medium (IGM) with metals and reveal the central accreting SMBH as an optically visible quasar (Silk \& Rees 1998; Moll et al. 2007). 
Several mechanisms have been proposed that could produce the force accelerating disk winds in AGN, including gas/thermal pressure (e.g., Weymann et al. 1982; Krolik \& Kriss 2001), magnetocentrifugal forces (e.g., Blandford \& Payne 1982; Everett 2005) and radiation pressure acting on spectral lines and the continuum (e.g., Murray et al. 1995, Proga et al. 2000). 
In reality, these three forces may co-exist and contribute to the dynamics of the outflows in AGNs to somewhat different degrees.

Observations of outflows are mainly carried out in absorption against the central compact UV/X-ray continuum. Absorption lines are usually classified based on the full width at half maximum (FWHM) of their profiles. The broad absorption lines (BALs) display deep, broad and typically smooth absorption troughs with velocity width larger than a few thousand $\rm km\, s^{-1}$. They clearly identify powerful outflows with typical velocities of order $\rm 20,000$ to $\rm 30,000\;km\,s^{-1}$, although velocities reaching $\rm 60,000\;km\,s^{-1}$ have also been measured \footnote{In this work the absorber velocity is assumed to be positive when the observed line is blue shifted with respect to the quasar emission redshift.} (Jannuzi et al. 1996; Hamann et al. 1997). BALs are detected almost exclusively in the most luminous radio-quiet quasars (RQQs) at a frequency of $\rm 12-23$ percent (Hamann et al. 1993; Hewett \& Foltz 2003, Ganguly et al. 2007). Relatively few radio-loud quasars (RLQs) are so far known to display BAL systems (i.e., Brotherton et al. 1998; Becker et al. 2001; Gregg et al. 2006).

The narrow absorption lines (NALs), on the other hand, have relatively sharp profiles with FWHM $\rm \lesssim 300 \;km\, s^{-1}$ so that important UV doublets, such as CIV$\rm\,\lambda\lambda\,1548,1550$ \AA, are discernible. Unlike their very broad kin, these absorbers are generally not blended and, therefore, offer a means to determine ionization levels and metallicities using absorption-line diagnostics. 
In addition, NALs are found with greater ubiquity, and detected (with varying frequency) in all AGN subclasses from Seyfert galaxies (e.g., Crenshaw et al. 1999) to higher luminosity quasars (e.g., Ganguly et al. 2001; Vestergaard 2003; Misawa et al. 2007) and from steep to flat radio spectrum sources (Ganguly et al. 2001; Vestergaard 2003), although the strongest ones appear preferentially in RLQ spectra (Foltz et al. 1986; Anderson et al. 1987).
Thus, NALs are more useful as probes of the physical conditions of outflows and provide an important tool to investigate the quasar environment. They may give us access to a different fraction, a different phase or different directions through the outflow with respect to BALs (e.g., Hamann \& Sabra 2004). 

The limitation of NALs is that they arise from a wide range of environments, from high speed outflows, to halo gas, to the unrelated gas or galaxies at large distances from the AGN. 
Since intrinsic gas clouds may also have small extents and masses similar to most intervening systems, below an absorption strength of about 1 \AA$\,$ we will get contaminations from intervening systems that are detected at similar redshifts (and thus velocities) as the quasar and that are difficult to identify as intervening.
The outflow/intrinsic origin of individual NALs can be inferred from: i) time-variable line-strengths, ii) resolved absorption profiles that are significantly broader and smoother compared with the thermal line widths, iii) excited-state absorption lines that require high gas densities or intense radiation fields, iv) line strength ratios in multiplets that reveal partial coverage of the background light source, v) higher ionization states than intervening absorbers. Nonetheless, NALs can still be connected to the quasar host without exhibiting such properties (Hamann et al. 1997).

Previous studies have shown that NALs tend to cluster near the emission redshift, at $\rm z_{abs}\approx z_{em}$.
A significant excess of absorbers over what is expected from randomly distributed intervening structures was measured by Weymann et al. (1979), with a distribution of intrinsic CIV NALs extending up to $\rm v\sim 18,000\;km\,s^{-1}$.  Foltz et al. (1986) confirmed such a statistical excess within $\rm \sim 5000\;km\,s^{-1}$ of $\rm z_{em}$.  
However, other studies failed to confirm any peak in the distribution of CIV absorption systems close to the emission redshift (Young et al. 1982; Sargent et al. 1988). 
After the work by Foltz et al. (1986), it has been traditionally assumed that NALs falling at less than $\rm 5000 \;km\, s^{-1}$  from the systemic redshift of the quasar are likely intrinsic and directly influenced by quasar radiation.
These systems have been commonly defined as $\it associated$ absorption lines (AALs) and those with velocity displacements greater than $\rm 5000\;km\,s^{-1}$, as $\it intervening$ systems. Despite the commonly adopted threshold of  $\rm 5000 \;km\, s^{-1}$ to distinguish intervening and associated absorbers, there is some evidence of a statistically significant excess of NALs extending to somewhat larger velocities (Misawa et al. 2007, Nestor et al. 2008, Tripp et al. 2008, Ganguly et al. 2013).

Large spectroscopic surveys provide a statistical means of measuring the frequency with which outflows are observed. This frequency can be easily related with the fraction of solid angle subtended by outflowing gas (e.g., Elvis 2000). Alternatively, the frequency can be interpreted as the fraction of the duty cycle over which AGNs produce outflows (assuming they subtend $\rm 4\pi$ sr). 

Ganguly \& Brotherton (2008) provide a review of the recent literature regarding the frequency of outflows, finding that, almost independently of luminosity, about 60 percent of AGNs show outflows in absorption (considering both BALs and NALs; see also Hamann et al. 2012).
However, this estimate is based on several simplifying assumptions and is affected by large uncertainties due to the inhomogeneity of the samples under consideration. In the case of NALs, the estimate of the frequency of outflows (based on the identification of intrinsic systems) has been carried out mainly in two ways: 1) with small samples of high redshift spectra, identifying intrinsic systems using mainly the partial coverage effect (Ganguly et al. 2001; Misawa et al. 2007; Ganguly et al. 2013). In particular, Ganguly et al. (2013) found a possible evidence of redshift evolution of the fraction of outflows with $9-19$ percent  at $\rm 0<z<0.7$, $14-29$ percent at $\rm 0.8<z<2$ and $43-54$ percent at $\rm 2<z<4$ (Misawa et al. 2007). 2) with very large samples at low resolution (based on SDSS) where the incidence of intrinsic systems is determined statistically by modeling the velocity offset distribution of absorbers (Nestor et al. 2008; Wild et al. 2008; Isee also Bowler et al. 2014). In particular, Nestor et al. (2008)  found a fraction of outflows of $\sim 14$ percent (for $\rm 0<v<12,000\; km\,s^{-1}$) which should be considered as a lower limit due to the uncertainties related with the low resolution and S/N of the spectroscopic sample. 

In this paper, we present a new survey of NALs to investigate the fraction of outflows in a previously unstudied redshift range, based on intermediate spectral-resolution observations from the echelle spectrograph X-shooter on the European Southern Observatory (ESO) Very Large Telescope (VLT) of 100 high redshift quasars ($\rm z_{em}\simeq 3.5-4.5$).
The target quasars were originally selected without regard to NAL properties, although BAL quasars were avoided. Thus, these data allow us to construct a large and likely unbiased sample of NALs. 
 L\'{o}pez et al. (2016; L\'{o}pez16, hereafter) provides details on the survey design.

The combination of high signal-to-noise ratio (S/N), wide wavelength coverage and moderate resolution of our survey have allowed us to look for empirical signatures to distinguish between the two classes of absorbers: intrinsic (produced in gas that is physically associated to the quasar) and intervening, without taking into account any a priori definition or velocity cut-off. 
We also take advantage of the large spectral coverage (from the UV cutoff at $\rm 300 \;nm$ to $\rm 2.5\; \mu m$) of our spectra to study the relative numbers of NALs in different transitions, indicative of the ionization structure of the absorbers and their locations relative to the continuum source.

The paper is organized as follows: $\S 2$ describes the properties of the quasar sample and briefly summarizes the observations; $\S 3$ describes our methodology for identifying NALs. Our results are presented in $\S 4$ and their implications in the context of relevant studies are discussed in $\S 5$. Our conclusions are summarized in $\S 6$. Throughout this manuscript, we adopt a $\rm \Lambda CDM$ cosmology with $\rm \Omega_M=0.315$, $\rm \Omega_{\Lambda}=0.685$, and $\rm H_0=67.3\;km\,s^{-1}Mpc^{-1}$ (Planck Collaboration et al. 2014).

\section{Quasar sample and observations}
\begin{figure}
   \centering
   \includegraphics[width=\hsize]{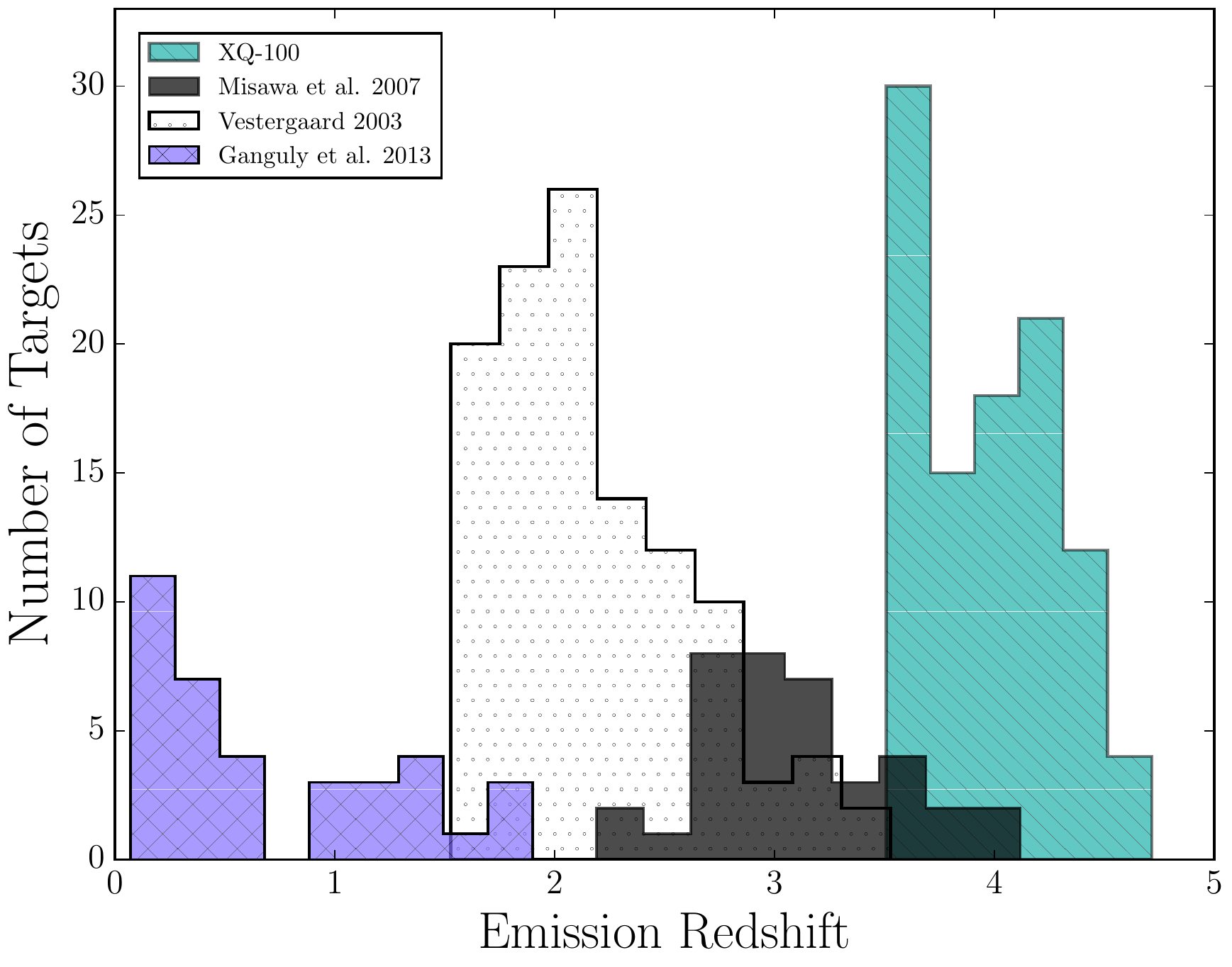}
     \caption{ Redshift distribution of our quasar sample (diagonal filled histogram), compared with previous works: Vestergaard 2003 (open dotted histogram); Misawa et al. 2007 (dark shaded histogram); Ganguly et al. 2013 (diamond filled histogram). XQ-100 enlarge the studied redshift range to higher values.}
       \label{FigVibStab}
  \end{figure}

The quasars in our sample have been originally selected and observed in a new Legacy Survey, hereafter "XQ-100", of 100 quasars at emission redshift $\rm z_{em}=3.5-4.5$ (ESO Large Programme 189.A-0424). The observations have been carried out with X-shooter/VLT (Vernet et al. 2011). The released spectra provide a complete coverage from the atmospheric cutoff to the NIR with a spectral resolution $\rm R\approx6000-9000$ depending on wavelength, and a median signal-to-noise ratio $\rm S/N\sim30$ at the continuum level. XQ-100 provides the first large intermediate-resolution sample of high-redshift quasars with simultaneous rest-frame UV/optical coverage. A full description of the target selection, observations, and data reduction process is presented by L\'{o}pez16. The distribution of quasar emission redshifts is shown in Fig.~\ref{FigVibStab} compared with those of other relevant works. 

XQ-100 was designed to address a wide range of high-redshift science (for an extensive list, see L\'{o}pez16). The targets were not selected on account of the presence of absorption lines in their spectra, although BAL quasars were avoided, so the sample is relatively unbiased with respect to the properties of NALs. Therefore, XQ-100 represents a unique dataset to study the rest-frame UV and optical of high-z quasars in a single, homogeneous and statistically significant sample.

The bolometric luminosities $\rm L_{bol}$  of all the quasars in our sample are estimated based on the observed flux at $\rm \lambda=1450$ \AA\,and using a bolometric correction factor $\rm L = 4.2 \;\lambda \;L_{\lambda}$ (Runnoe et al. 2012).

An important point for our study is to obtain accurate systemic redshift for the quasars in order to determine the relative velocity/location of the absorbers. 
This is crucial especially for the associated systems. The quasar emission redshifts of our sample have been obtained from a Principal Components Analysis (PCA; see e.g., Suzuki et al. 2005). For a complete description of the procedure, see L\'{o}pez16.

\subsection{Radio properties}
Previous studies have shown that there is a possible dependence of the presence of associated systems on the radio properties of the inspected quasars (Baker et al. 2002, Vestergaard 2003). For this reason, we have also investigated the radio properties of the XQ-100 sample.

Matching our sample with the FIRST catalogue (Becker et al. 1995), we have found radio information for 67 over 100 quasars. According to our determination of the radio-loudness parameter, 12 objects are radio-loud and 55 are radio-quiet. Most likely the objects which are not in the FIRST catalogue are also radio-quiet. As a consequence, we will consider the XQ-100 sample as a substantially radio-quiet sample. 

The radio-loudness of a quasar is typically parametrized by the ratio between the rest-frame flux densities at $\rm 5\;GHz$ and $\rm 2500$ \AA, i.e., 
$\rm R = f_{\nu}(5\; GHz)/f_{\nu}(2500)$ \AA\,(Sramek \& Weedman 1980). 
We compute the radio flux density at rest-frame $\rm 5\;GHz$, $\rm f_{\nu}^{rest}(\rm 5\;GHz)$, from the observed flux density, $\rm f_{\nu}^{obs}(\nu)$, at observed frequency $\rm \nu$:

\begin{equation}
\begin{split}
\ \rm log\,f_{\nu}^{rest}(\rm 5\;GHz) &= \rm log\,f_{\nu}^{obs}(\nu)+\alpha_r log(5GHz/\nu)\\
  &\rm -(1+\alpha_r)\;log\,(1+z_{em}),
\end{split}
\end{equation}

\begin{raggedleft}{
 {where $\rm \alpha_r$ is the spectral index, $\rm f_{\nu}\sim\nu^{\alpha}$. We assume $\rm \alpha_r=-0.5$ (e.g., Ivezic et
al. 2004).} The flux density $\rm f_{\nu}^{rest}(\rm 2500)$ \AA, is derived from a power-law fit to our own data.
}
\end{raggedleft}

Following Ganguly et al. (2013), we adopt $\rm R\geq23$ as the criterion for radio loudness.

The most important properties of the quasars in our sample are summarized in the Table~\ref{tab_ioni}.  Columns (1) and (2) of the Table~\ref{tab_ioni} give the quasar name and emission redshift, column (3) the bolometric luminosity and column (4) quasar radio type.

\begin{table*}
\caption{The full table is available online.}
\begin{tabular}{c c c c c  c c c c c }
  \hline
  \hline
Quasar & $\rm z_{EM}$  &$\rm log\, L_{bol} $ &Radio type &$\rm z_{abs} $ & ion  & $\rm log\, N_a$ &$\rm W_0$ \\ 
\small (1) &\small (2) &\small(3) & \small(4) &\small (5)  &\small(6) &\small(7)&\small (8)\\ 
  \hline
J1013+0650    &     3.808& $46.100\pm  0.014$& RL & 2.911&CIV&$ 13.92\pm  0.04 $ &$0.278  \pm0.007$ \\
&  & &  & 2.944 &CIV & $13.39\pm 0.05$& $0.091 \pm 0.006$ \\
&  &  & & 3.081 &CIV & $13.48\pm 0.05$& $0.098 \pm 0.005$ \\
&  &   && 3.236 &CIV & $14.14\pm 0.06$& $0.428 \pm 0.005$ \\
&  &  & &  &SiIV & $13.56\pm 0.04$& $0.266 \pm 0.005$ \\
&  &  & & 3.283 &CIV & $13.95\pm 0.04$& $0.318 \pm 0.006$ \\
&  &  & &  &SiIV & $12.68\pm 0.03$& $0.046 \pm 0.005$ \\

\hline
\end{tabular}
\label{tab_ioni}
\end{table*}

\section{Sample of NAL systems}

\subsection{Identification and measurement of CIV absorbers}

With the aim of mapping the incidence of NALs in quasar rest-frame velocity space, we produce a catalog of the ultraviolet doublet CIV $\rm \lambda \lambda$ 1548.204, 1550.78\footnote{ Wavelengths and oscillator strengths used in this work are adopted from Morton (2003). }\AA.\, Since we do not impose an a priori velocity definition of associated system, we look for any CIV absorber outside the Ly$-\alpha$ forest in each spectrum.

Although continuum estimates are available for the XQ-100 public data release, we elect to perform our own continuum fits, due to the sensitive nature of accurate continuum placement close to the quasar emission lines. For each target the continuum level is determined by fitting with a cubic spline the portions of the spectrum free from evident absorption features in the region redward of the Ly$-\alpha$ emission of the quasar. We then normalize to the continuum level the flux and error arrays.
We visually inspect all the quasar spectra looking for CIV doublets using the LYMAN context of the MIDAS reduction package (Fontana \& Ballester 1995). 
We select all the candidates $\rm \lambda 1548$ and $\rm \lambda 1550$ transitions with matching kinematic profiles. 
We make use of the TELLURIC task of the Image Reduction and Analysis Facility package (IRAF; Tody 1986) to apply appropriate corrections to each spectrum in the spectral regions affected by the telluric bands. This latter step allows us to remove telluric features and to have a more firm identification of the lines.
At a first run through, absorption troughs that are separated by unabsorbed regions are considered to be separate lines. In this manner, 1098 CIV doublets are identified. 

The rest-frame equivalent width ($\rm W_0$) and its measurement error are measured for each line by integrating across the flux density and error array, respectively, over a user-defined interval. The interval extremes are defined by the wavelengths where the flux matches again the level of the continuum. For the statistical analysis we include in our sample of NALs only doublet lines whose weaker member is detected at a confidence level greater than $3\sigma$ i.e., each $\rm W_{1550}$ is larger than the detection limit, given by:

\begin{equation}
\\\\\ \rm{W_{lim}}=\rm {n_{\sigma} \frac{FWHM}{(1+z_{abs})\; S/N}}
\label{W_tresh}
\end{equation}

\begin{raggedleft}{
where, $\rm {n_{\sigma}}$ is equal to 3, $\rm FWHM$ is computed as the ratio of the observed wavelength $\lambda$ and the spectral resolution, $\rm z_{abs}$ 
is the absorption redshift of the line and $\rm S/N$ is the signal to noise ratio per pixel in the consider region.
}
\end{raggedleft}

The NAL sample defined by this limit contains 1075 CIV doublets.
Furthermore, as CIV absorption is relatively common in quasar spectra, blending of systems at similar redshifts could be a problem. Therefore, when either the $\rm \lambda 1548$ or the $\rm \lambda 1550$ CIV component is blended with other transitions we include the doublet in the complete sample only if the doublet ratio, defined as the ratio of equivalent widths of the stronger to the weaker component, is in the range 0.8-2.2. The theoretical ratio for this doublet is 2, so the above mentioned interval is adopted to account for blending effects and is determined from CIV narrow absorption doublets  clearly not affected by blending. Adopting this criterion the sample is reduced to 1060 doublets. 

To refine our sample for statistical analysis, we combine NALs that lie within  $300\;\rm km\,s^{-1}$ of each other into a single $\it system$. This makes the sample homogeneous. This clustering velocity is chosen considering the largest velocity extent of an absorber in our sample for which it is not possible to discern the individual components. Therefore, we take $300\;\rm km\,s^{-1}$ as the minimum velocity separation for which two absorbers are counted separately.
To identify CIV systems, we proceed in the following way: for each list of CIV components corresponding to a single quasar the velocity separations among all the lines are computed and sorted in ascending order. If the smallest separation is less than $\rm dv_{min}=300\;\rm km\,s^{-1}$ the two lines are merged into a new line with equivalent width equal to the sum of the equivalent widths, and redshift equal to the average of the redshifts weighted with the equivalent widths of the components. The velocity separations are then computed again and the procedure is iterated until the smallest separation becomes larger than $\rm dv_{min}$.

The absorber velocity with respect to the quasar systemic redshifts is conventionally defined as $\rm v_{abs} = \beta\, c$, we compute it by the relativistic Doppler formula (e.g., see Vestergaard 2003),

\begin{equation}
\\\\\\\ \rm \beta \equiv \frac{v_{abs}}{c} =  \frac{(1+z_{em})^2 - (1+z_{abs})^2}{(1+z_{em})^2 + (1+z_{abs})^2}
\label{v_abs}
\end{equation}

\begin{raggedleft}{where $\rm z_{em}$ and $\rm z_{abs}$ are the emission redshift of the quasar and the absorption redshift of the line, respectively and $\rm c$ is the speed of light.}\end{raggedleft}
Our final sample consists of 986 CIV doublets with $\rm -1000 < v_{abs} < 73,000\;\rm km\, s^{-1}$ and equivalent widths 0.015 \AA\,$\rm< W_0 < $2.00\,\AA. 

We finally measure the column densities ($\rm N$) of the lines with the apparent optical depth (AOD) method (Savage \& Sembach 1991;  Sembach \& Savage 1992). This method provides a quick and convenient way to convert velocity-resolved flux profiles into reliable column density measurements for intergalactic absorption lines, without the need to follow a full curve-of-growth analysis or detailed component fit and without requiring prior knowledge of the component structure. The latter aspect is particularly important when high resolution data are not available. 
In the AOD method, a velocity-resolved flux profile $\rm F(v)$ is converted to an apparent optical depth profile $\rm \tau_a(v)$ using the relation

\begin{equation}
\\\\\ \rm  \tau_a(v)=ln [ F_c(v)/F(v) ],
\end{equation}

\begin{raggedleft}{where $\rm F_c(v)$ and $\rm F(v)$ are the continuum and the observed line fluxes at velocity $\rm v$, respectively. Then the apparent optical depth profile can be converted into an apparent column density profile according to}\end{raggedleft}

\begin{equation}
\\\\\ \rm N_a(v)=3.768\times 10^{14}\,(f\lambda)^{-1}\,\tau_a(v)\;cm^{-2}\,(km\,s^{-1})^{-1},
\end{equation}

\begin{raggedleft}{where $\rm f$ is the oscillator strength of the transition and $\rm \lambda$ is the transition restframe wavelength in \AA\, (Savage \& Sembach 1991). The total apparent column density is then }\end{raggedleft}

\begin{equation}
\\\\\ \rm N_a=\int_{v_-}^{v^+} N_a(v)\,dv
\end{equation}

\begin{raggedleft}{where $\rm v_-$ and $\rm v^+$ are the velocity limits of the line. This technique offers very good results if applied to spectra with S/N$\gtrsim 20$ (Fox et al. 2005). When two lines of different strength of the same ionic species are available, the AOD method allows us to assess and correct for the level of saturation in the data by comparing the apparent column density derived from the stronger line with that derived from the weaker one (Savage \& Sembach 1991; Jenkins 1996). Therefore, we take advantage of this method to correct the column density of the CIV doublets when the stronger component is saturated.}\end{raggedleft}  We define as saturated lines which have flux density levels at the wavelength of peak absorption of less than 0.2 in the normalized spectra (Krogager et al. 2016). When both components are saturated we do not compute the column density for the system in question.

\subsection{Identification of other species}
The detection of a single species in a single ionization state substantially limits the information on the nature and the physical properties of the studied absorbers. The quality and wavelength extent of our spectra give us the possibility to search for other common ions (NV, SiIV and CII) related to each detected CIV absorber. Furthermore, for the ions mentioned above, we search for absorption lines not related to the identified CIV, but we do not find any.

We identify NV, SiIV doublets and CII in the following regions:
\begin{enumerate}
\item NV $\rm \lambda \lambda 1238.821, 1242.804$ \AA\,\,absorption doublets: from $-1000$ to $5500\; \rm km\, s^{-1}$ with respect to the quasar's NV emission line. The velocity range is relatively narrow for this transition because contamination by the $\rm Ly\alpha$ forest prevents us from searching for NV NALs at larger velocity offsets;
\item SiIV $\rm \lambda \lambda 1393.760, 1402.772$ \AA\,\,absorption doublets: from  $-1000$ to $45,000 \;\rm km\, s^{-1}$ with respect to the quasar's SiIV emission line;
\item CII $\rm \lambda 1334.532$ \AA\,\,absorption: from $-1000$ to $10,000 \;\rm km \,s^{-1}$ with respect to the quasar's CII emission line.
\end{enumerate}

\begin{table}
\caption{Number and fraction of CIV systems with detected NV, SiIV and CII.}
\begin{minipage}{80mm}
\label{tab_ions}
\begin{tabular}{c c  c c }
\hline
\hline
 {}&\scriptsize{$ \rm -0.1<v_{abs}^a<0.25$}  & \scriptsize{$\rm 0.25<v_{abs}^a<0.55$}&  \scriptsize{$\rm v_{abs}^a<0.5$}\\
 \hline
\small{CIV} &$72$& 68 &$122$\\
\hline
 NV &$34; 47\%$ &$12; 18\%$&$43; 35\%$\\

SiIV & $26; 36\%$&$27; 40\%$&$46; 53\%$\\

 CII&$7; 10\%$&$10; 15\%$&$16; 13\%$\\

\\
\end{tabular}

$^a$ $\rm v_{abs}$ is measured in unity of $\rm 10^4\;km\,s^{-1}$.\\ {\bf Note:} the bin in the second column is extended to $\rm 5500\;km\,s^{-1}$ to include all the NV absorptions detected. The third column shows the number and fraction of associated CIV systems with other detected ions. 
\end{minipage}
\end{table}

The same procedures of analysis described above for CIV are applied to NV, SiIV and CII as well. In particular, we require the detection level for the weaker member of the doublets to be 3$\sigma$,  and 5$\sigma$ for single absorption lines.
The final sample includes 574 SiIV (234 detections, 340 upper limits), 140 NV (46 detections, 94 upper limits) and 142 CII (28 detections, 114 upper limits). See Table~\ref{tab_ions} for more details on the associated region. The non-detections are reported as upper limits. They are calculated by integrating across the expected location of each species in the spectrum, at a velocity extension similar to that of the corresponding CIV absorber. This procedure is actuated when spurious absorption lines are not present in the considered portion of the spectra.

For the remainder of the paper, $\rm W_0$ represents the rest frame equivalent width of the stronger member of a given doublet. Columns (7) and (8) of Table~\ref{tab_ioni} report the apparent column density of the absorbers and the equivalent width of the lines combined into single $\it systems$ within $\rm 300\;km\,s^{-1}$, respectively. Examples of the identified absorbers are reported in the appendix.

\subsection{Completeness Limits}

\begin{table*}
\caption{Results of the completeness test for various absorber velocities, parametrized by $\rm \beta$. }             
\centering          
\begin{tabular}{c c c c c c c c }     
\hline\hline       
\\
\multicolumn{8}{c}{Fraction of quasars with $\rm W_{lim}$ [\AA]}\\ 
\\
\hline
  &$\scriptstyle{ \rm W_{lim}< 0.005\,}$    & $\scriptstyle{  \rm W_{lim}>0.005\,}$ & $\scriptstyle{ \rm W_{lim}>0.01\,}$&$\scriptstyle{ \rm W_{lim}>0.015\,}$&$\scriptstyle{ \rm W_{lim}>0.02\,}$&$\scriptstyle{ \rm W_{lim}>0.025\,}$&$\scriptstyle{ \rm W_{lim}>0.03\,}$\\

\hline                    
   $\rm \beta=0.00$  & 0.31 & 0.84 & 0.97 & 0.99 & 1.00 & 1.00 & 1.00\\      
   $\rm \beta=0.01$  & 0.21 & 0.69 & 0.90 & 0.97 & 1.00 & 1.00 & 1.00\\
   $\rm \beta=0.02$  & 0.12 & 0.57 & 0.86 & 0.93 & 0.98 & 1.00 & 1.00\\
   $\rm \beta=0.03$  & 0.08 & 0.49 & 0.79 & 0.91 & 0.96 & 0.99 & 1.00\\
   $\rm \beta=0.04$  & 0.07 & 0.42 & 0.77 & 0.89 & 0.97 & 0.99 & 0.99\\ 
   $\rm \beta=0.10$  & 0.07 & 0.39 & 0.76 & 0.90 & 0.97 & 1.00 & 1.00\\ 
\hline                  
\end{tabular}
\label{tab_test_comp}
\end{table*}

 \begin{figure}
   \centering
   \includegraphics[width=8.3cm]{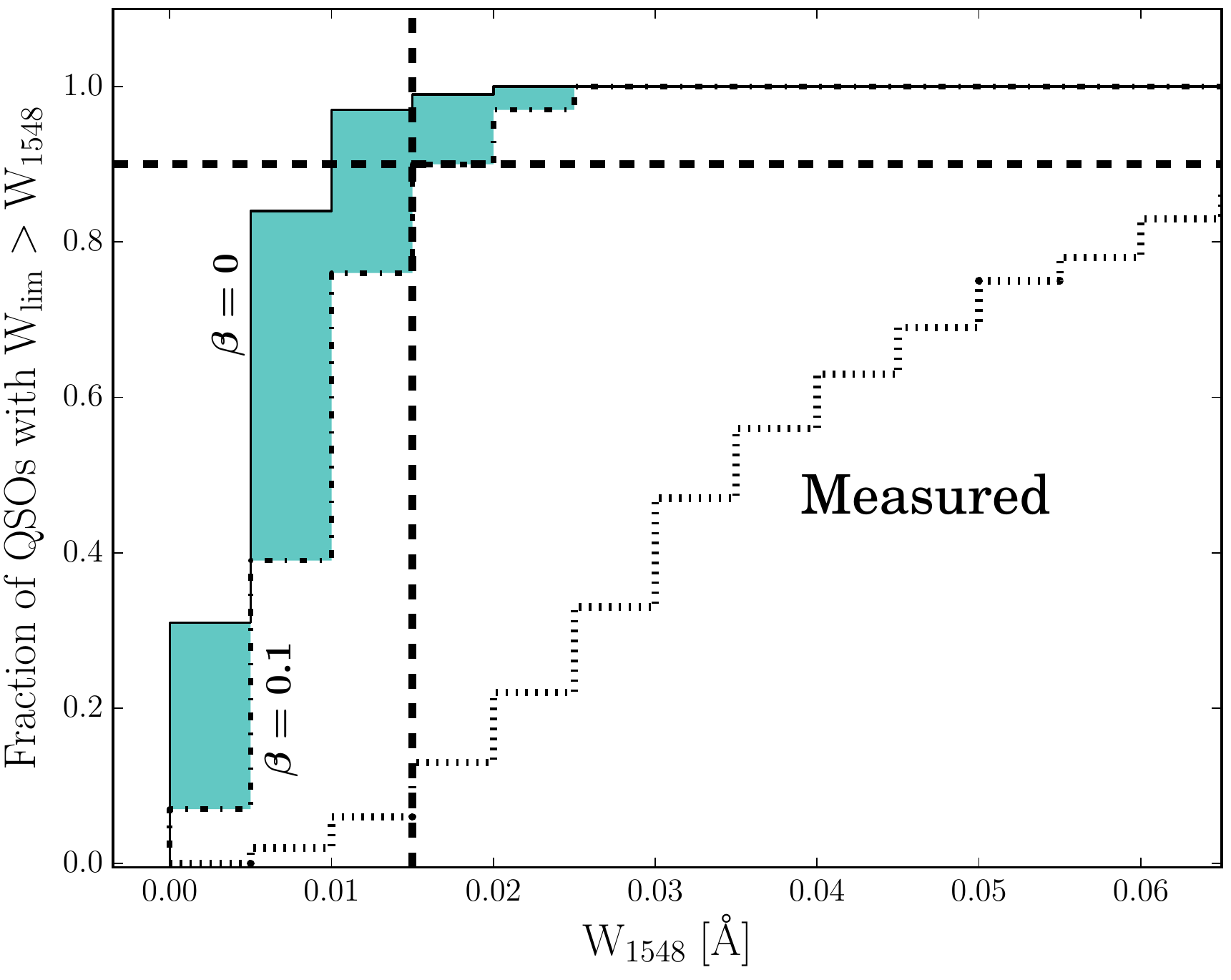}
     \caption{Cumulative distribution of the number of quasars with $\rm 3\,\sigma $ rest-frame $\rm W_{lim}$ detection limit larger than a given $\rm W_{1548}$ value and for several $\rm \beta$-values. The black solid line represents the distribution for $\rm \beta=0.0$, the dot-dashed one represents the $\rm \beta=0.1$ distribution. The cumulative distributions for $\rm \beta > 0.1$ overlap with the $\rm \beta=0.1$ distribution and are not marked. The cumulative distribution of the number of quasars showing at least one detected absorber with $\rm W_{1548}$ larger than a given threshold is reported for comparison (dotted line, labeled "Measured"). The 90 percent completeness level is marked by the dashed horizontal line. The $\rm 3\,\sigma$ detection limit measured to a completeness level of $\sim90$ percent is $\rm W_{1548}=0.015$ \AA\,(dashed vertical line).
             }
  \label{test_comp}  
  \end{figure}
  
In order to estimate the completeness of our sample, we test the detection sensitivity of the whole dataset. We compute the cumulative distributions of the number of quasars with $\rm W_{lim}$ larger than a given value of $\rm W_{1548}$ and within a given velocity separation from the quasar emission redshift. $\rm W_{lim}$ is measured from Eq.~\ref{W_tresh}, for various absorber velocities (parametrized by $\rm \beta$) in the spectra of all quasars. The results are shown in Fig.~\ref{test_comp}. 

The black solid line and the dot-dashed one represent the distributions for $\rm \beta=0.0$ and $\rm \beta=0.1$, respectively. The lines corresponding to the cumulative distributions relative to values of $\rm \beta$ between $\rm \beta=0$ and $\rm \beta=0.1$ span the shaded area in the plot.  For increasing $\rm \beta$-values, the cumulative distributions overlap the one with $\rm \beta =0.1$. 

The cumulative distributions show that in our sample the CIV detection sensitivity drops below 90 percent at $\rm W_{1548}\approx 0.015$ \AA\,for $\rm \beta \geq 0.1$ (vertical and horizontal dashed lines in Fig.~\ref{test_comp}). This is explained by the identical spectral coverage and the uniformity in S/N of the spectra in our sample.
More explicitly, at $\rm \beta=0.02 \;(0.10)$, corresponding to $\rm 6000\;km\,s^{-1}$ ($\rm 30000\;km\,s^{-1}$), the completeness level of $\rm W_{lim}\geq 0.015 $ \AA\,absorbers is $\geq 93$ percent $ \;(\geq 90$ percent).  
Part of the cumulative distribution of the number of quasars with at least one identified CIV line stronger than a given threshold is also shown in Fig. ~\ref{test_comp} for comparison (dotted line). It is noticeable from the plot that more than 87 percent of the quasars in our sample have at least one CIV detection with $\rm W_{1548}$ above the 0.015 \AA\,completeness limit. 
The results of the completeness test are collected in Table~\ref{tab_test_comp}.

\section{Results}

Using the complete CIV sample described in the previous section, we first investigate the equivalent width distribution and the number density of absorbers per unit of velocity interval as a function of the velocity separation from the quasar, i.e., the velocity offset distribution, $\rm dn/d\beta = c \,dn/dv$, where $\rm \beta$ is computed in Eq.~\ref{v_abs}. Then, we examine the relative number of NALs in different transitions and their velocity offset distributions. The goal is to investigate the ionization structure of the absorbers and their locations relative to the continuum source. 
We also estimate the number of intrinsic NALs per quasar, and the fraction of quasars hosting intrinsic NALs. These values can be used to study the geometry of absorbing gas around the quasar. 

To further characterize the metal absorptions associated to galaxies hosting $\rm z\sim4$ quasars, we examine the distributions of the covering fractions. 
Finally, we briefly look for dependencies of the NAL properties relative to those of their quasar hosts.
We choose to show the plots of this section in equivalent width and not in column density to better compare our results with previous works present in the literature. The statistical significance of the results remains the same using column density. 

\subsection{Equivalent Width Distribution}
\begin{figure}
   \centering
   \includegraphics[width=\hsize]{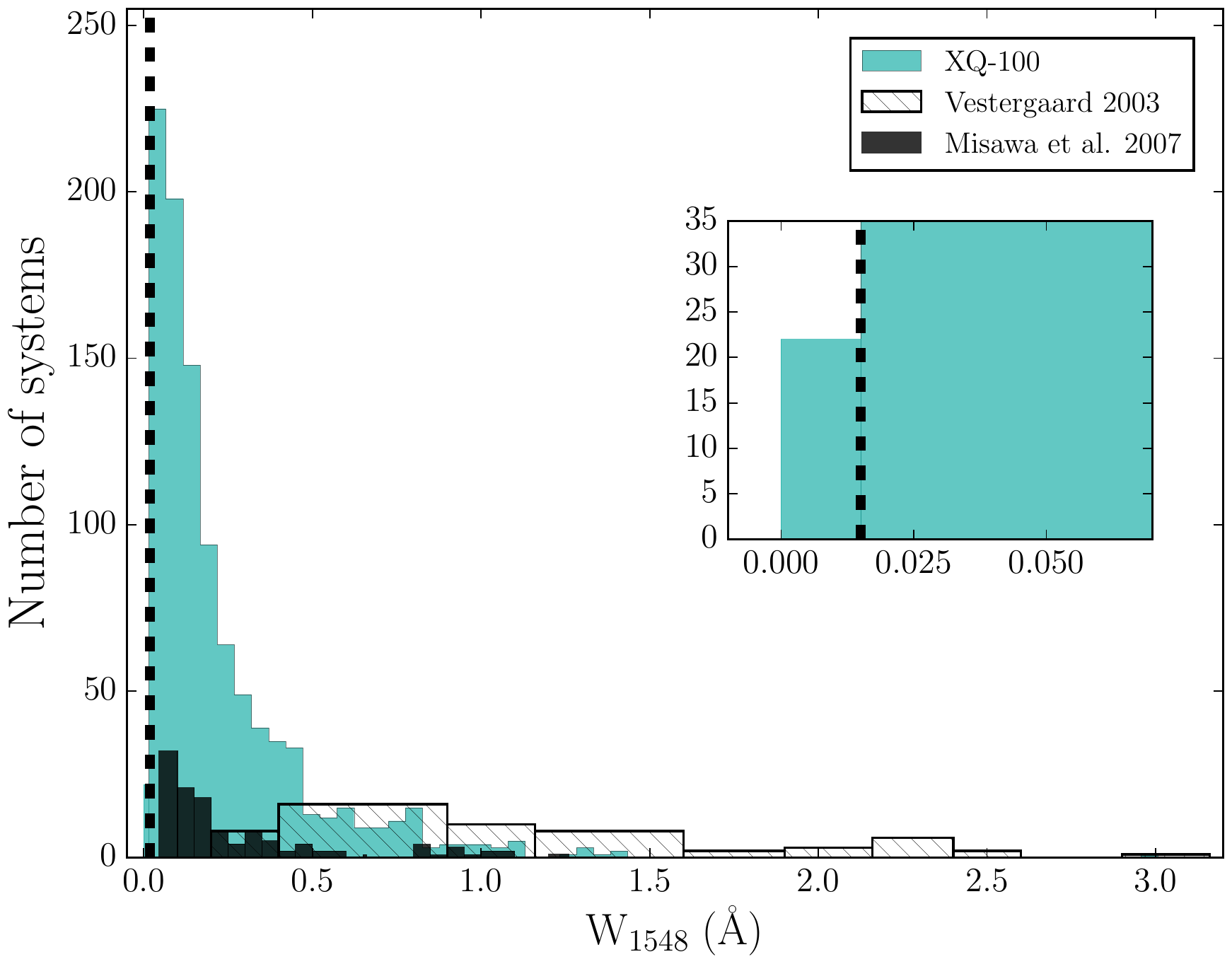}
     \caption{The rest-frame equivalent width distribution of all CIV systems in our sample (light shaded histogram). The vertical dashed line marks our detection limit of 0.015 \AA. Distributions measured by previous works are also shown: Vestergaard 2003 (diagonal filled histogram) and Misawa et al. 2007 (dark shaded histogram). The insert shows the details of the bottom left region of the histogram.
             }
             \label{ew_dist}
  \end{figure}

The distribution of rest-frame equivalent widths of the CIV absorbers is shown in Figure~\ref{ew_dist} (light shaded histogram). It rises steeply towards the detection limit of 0.015 \AA, marked in the diagram by the dashed line.

It is evident from the insert of Figure~\ref{ew_dist} that very few of the identified lines
 are excluded from the analysis because are weaker than the detection level required.
 We can also note that more than half of the absorbers ($\sim 63$ percent) have $\rm W_{1548} < 0.2$ \AA. Thanks to the intermediate resolution of X-shooter and to the high S/N level of the XQ-100 sample, we are probing much weaker lines than previous works at lower spectral resolution. For example, in the study of Weyman et al. (1979) the rest-frame equivalent width limit was about 0.6 \AA, while Young et al. (1982), Foltz et al. (1986) and Vestergaard (2003) reach a limit of about 0.3 \AA, larger than 91 percent and 76 percent of the CIV lines in our sample, respectively. 
Also Misawa et al. (2007), a study based on data at larger resolution than ours, obtain a detection limit of $\sim 0.056$  \AA, larger than 17 percent of NALs detected in our sample, due to the lower S/N of their spectra. 

 Equivalent width distributions from Vestergaard (2003, diagonal filled histogram) and Misawa et al. (2007, dark shaded histogram) are also shown in Fig.~\ref{ew_dist} for comparison.  The XQ-100 survey data can better detect and resolve weak lines with respect to Vestergaard (2003). The comparison with Misawa et al. (2007) illustrates that high resolution data, able to detect and better resolve weak lines, are usually based on less statistically significant samples.

\subsubsection{Dependence on radio properties}

We investigate possible differences between RLQs and RQQs with respect to the NAL properties and we do not find any.
A direct comparison of the incidence of the absorbers with previous works is non-trivial.
The aspects that play a crucial role are the spectral resolution and the different properties of the targets. Two different samples can have: i) differences in the radio properties and intrinsic luminosities of the targets, ii) different ranges in $\rm W_{1548}$ of the detected absorbers, iii) different number of targets with a full coverage of the associated region. 

With the aim of comparing the equivalent width distribution in our analysis with previous studies, we focus on systems with $\rm W_{1548} > 1.5$ \AA\,following Vestergaard (2003) who claimed the strongest systems are likely intrinsic candidates. The Vestergaard sample has 66 RLQs and 48 RQQs, nearly 
the same number, unlike ours, in which only 17 percent (12/67) of the targets for which we have information about radio properties, are radio-loud. If we consider separately the radio-loud and radio-quiet objects in the Vestergaard sample, only 5/66 and 1/48 quasars exhibit at least one CIV NAL with $\rm W_{1548} > 1.5$ \AA, respectively. Thus, a weighted prediction based on the fraction of RQQ and RLQ in our sample on the number of very strong CIV NALs is that two should be detected. Indeed, we do detect exactly two such absorbers with $\rm W_{1548}  > 1.5$ \AA. 
The fraction of radio-loud quasars in our sample (17 percent) is only slightly larger than the general population $(\sim10$ percent). As a consequence, we do not expect to be biased toward strong NALs in RLQs, unlike Vestergaard (2003) whose sample was formed by more than 50\% RLQs since the aim was to compare the NALs in RLQs and RQQs. The distribution of weaker lines is considered in the next sections.

\subsection{Velocity Offset Distributions}

  \begin{figure*}
  \centering
  \includegraphics[width=8cm]{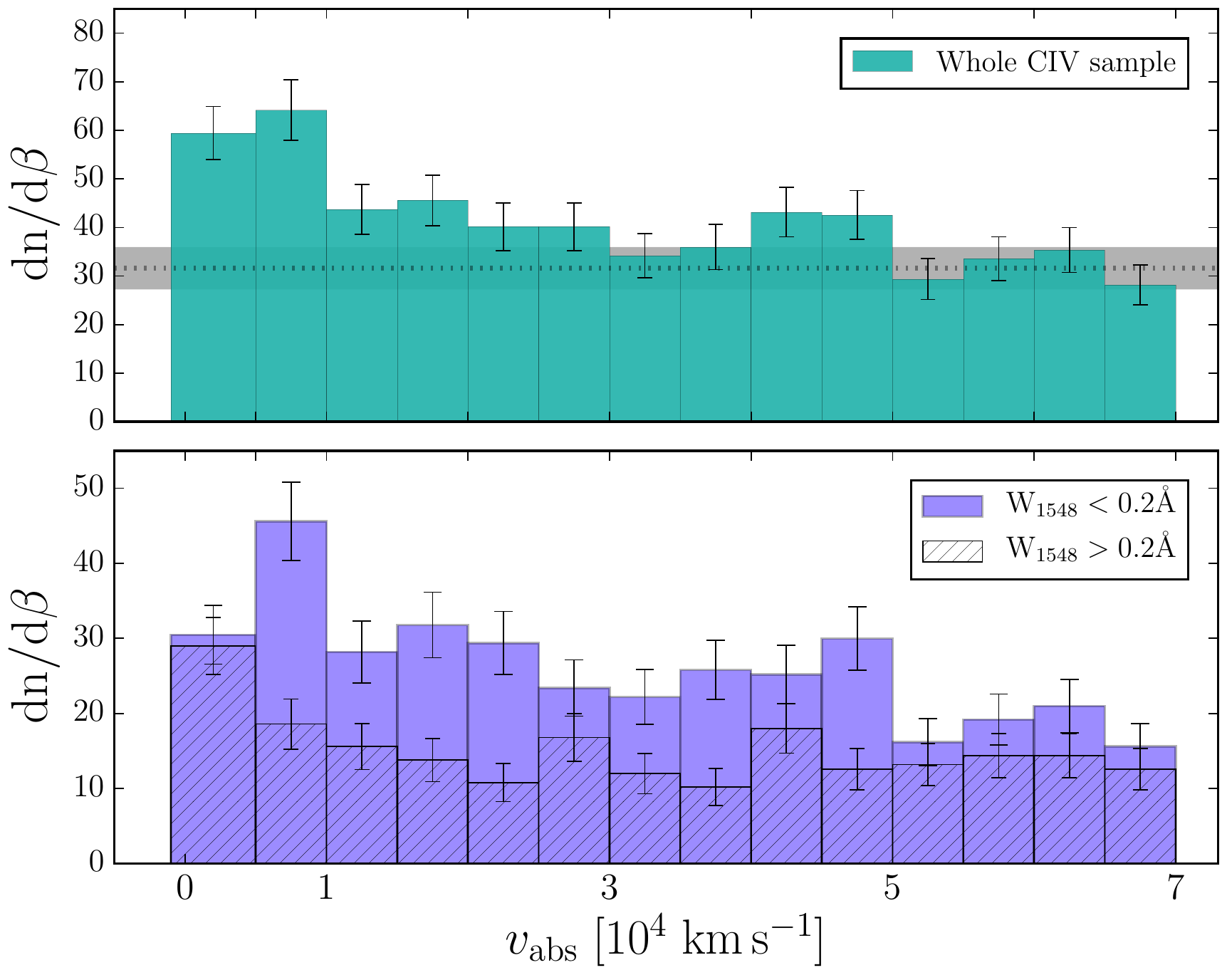}
  \qquad\qquad
  \includegraphics[width=8cm]{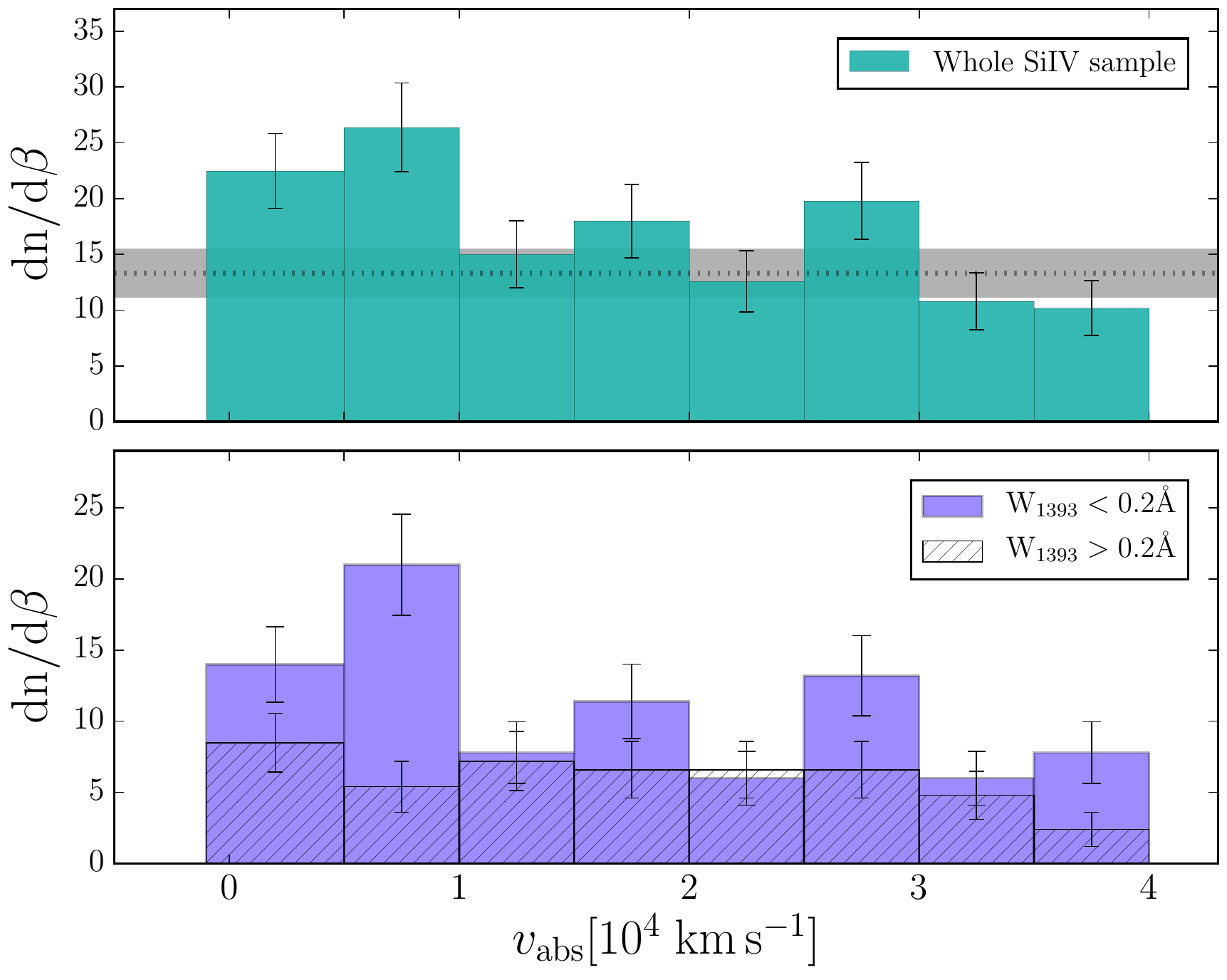}
  \caption{Velocity offset distribution of the number density of CIV (left) and SiIV (right) absorbers. Top panels: whole sample. Bottom panels: the color-shaded histogram represents the distribution of NALs with $\rm W_{0} <0.2$ \AA, the black hashed histogram represents the distribution of NALs with $\rm W_{0} >0.2$ \AA. The error bars represent the propagation of the Poissonian uncertainties. The horizontal dotted line represents the average number of systems measured in bins of $\rm 5000\;km\,s^{-1}$ far from the $\rm z_{em}$, and the shaded area around it is the $\rm 1\sigma$ error of this mean value. Contamination by the Ly-$\rm \alpha$ forest prevented us from exploring the same velocity range for the two ions.} 
  \label{offset_vel}
\end{figure*}

  \begin{figure*}
  \centering
  \includegraphics[width=8cm]{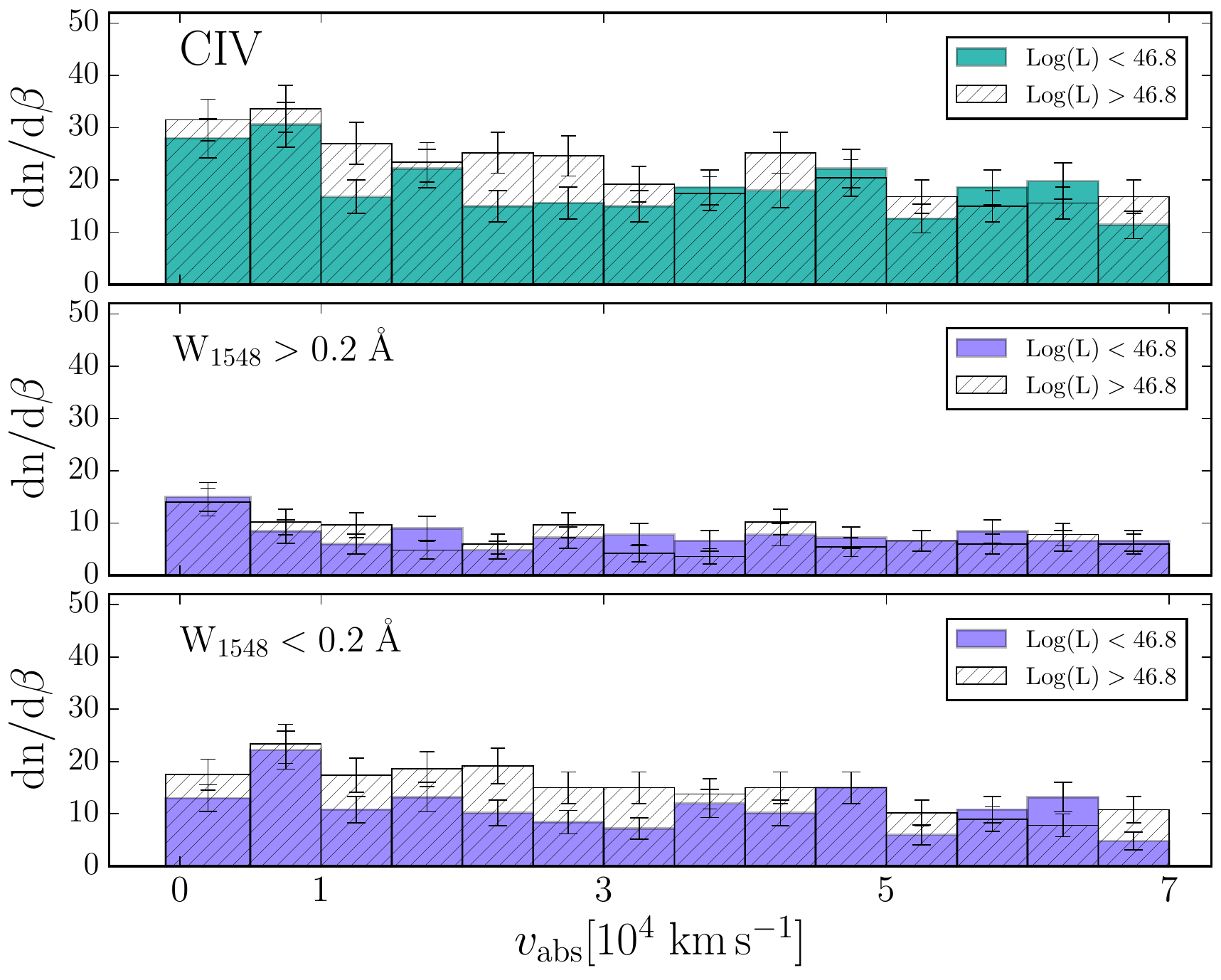}
  \qquad\qquad
  \includegraphics[width=8cm]{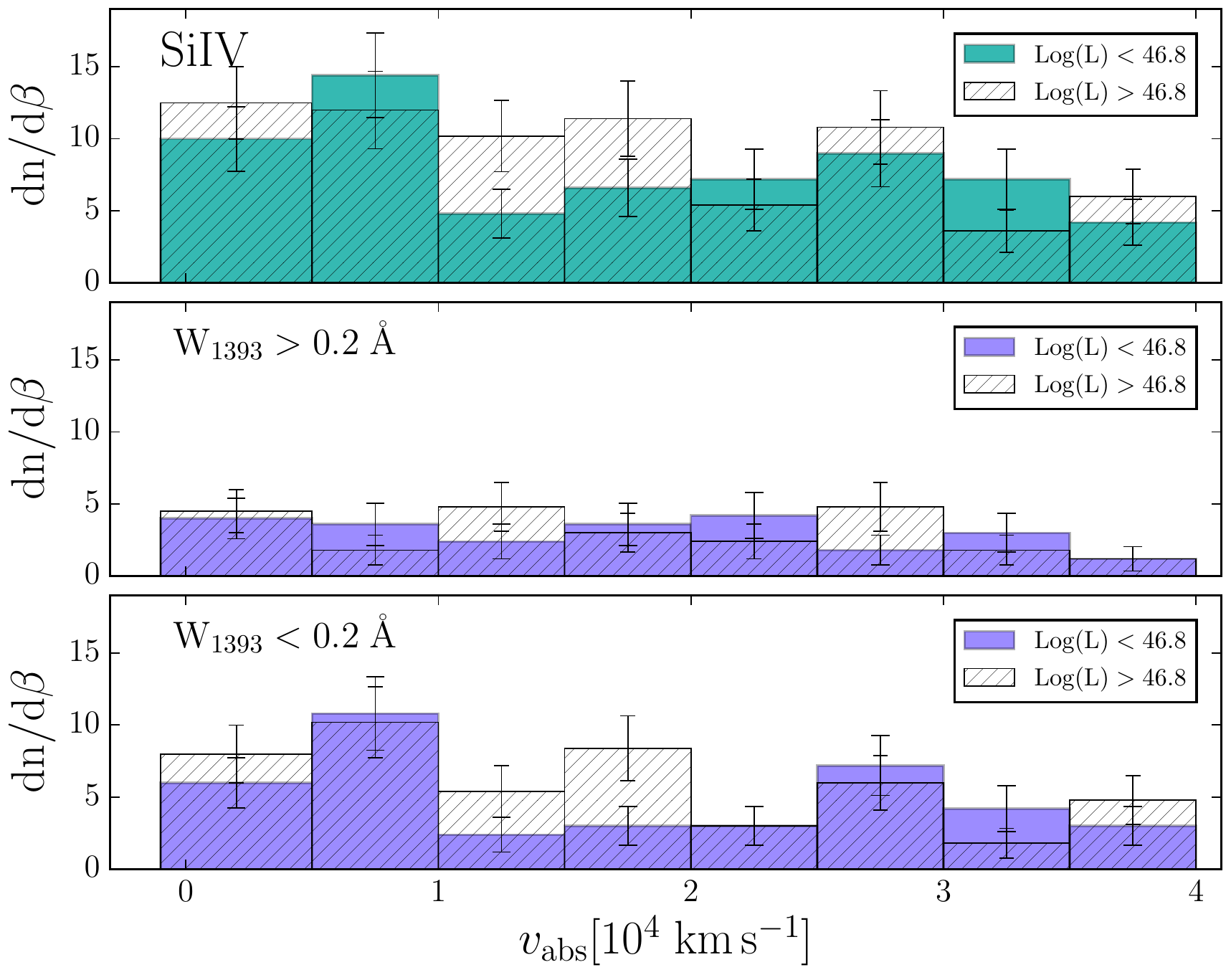}
  \caption{Velocity offset distribution of the number density of CIV (left) and SiIV (right) absorbers. The color-shaded histogram represents the distribution of NALs with $\rm L_{bol}<46.8\; erg\;s^{-1}$, the black hashed histogram represents the distribution of NALs with $\rm L_{bol}>46.8\; erg\;s^{-1}$. Top panels: whole sample. Middle panel: subsample of lines with $\rm W_{0}>0.2$ \AA. Bottom panels:  subsample of lines with $\rm W_{0}<0.2$ \AA. The error bars represent the propagation of the Poissonian uncertainties.} 
  \label{offset_vel_L}
\end{figure*}

The velocity offset distributions of the number density of CIV and SiIV NALs in our study are presented in Fig.~\ref{offset_vel}, left and right panels, respectively. 
In both diagrams the upper panel shows the distribution of the whole sample while in the bottom one the sample is split in two according to the strength of the NALs' equivalent widths: the color-shaded histogram represents the weak lines ($\rm W_0  < 0.2$ \AA\,) and the black hashed histogram the strong lines ($\rm W_0  > 0.2$ \AA\,). Error bars indicate $1\,\sigma$ Poissonian uncertainties. The zero velocity corresponds to the emission redshift of the quasar. 
The observed distributions include contributions from: i) intervening systems; ii) environmental absorption which arises either in the interstellar medium of the AGN host galaxy or in the IGM of the galaxy's group or cluster; and iii) outflow systems that are ejected from the central AGN. 

We do find an excess of the number density of NALs within $\rm 10,000\;km\,s^{-1}$ of the emission redshift. The excess is clearly detected in both the velocity offset distributions of CIV and SiIV NALs within the first two bins, starting from $\rm v_{abs}=-1000\;km\,s^{-1}$.
Since there is not a sizeable change in sensitivity to finding absorbers as a function of velocity, as shown in $\S \,3.2$,  and our spectra have all the same wavelength coverage, beyond the first two bins we see an approximately constant value expected for intervening systems. 

To assess whether the detected excess is statistically significant over random incidence or not, we firstly compute the average occurrence of CIV absorption in bins of $\rm 5000 \;km\,s^{-1}$ at large velocity separation ($\rm v_{abs}>5\times 10^4\;km\,s^{-1}$) from $\rm z_{em}$, where we are more confident that systems are intervenings. The average number density of CIV expected for the whole sample is $31\pm4$ per $\rm dv=5000\; km\;s^{-1}$ bin, to be compared with $\rm dn/d\beta=59$ and 64 that we find at velocity separations of $\rm -1000 <v_{abs}< 5000\; km\,s^{-1}$ and $\rm 5000 <v_{abs}< 10,000\; km\,s^{-1}$, respectively. This corresponds to a $\rm \sim8\,\sigma$ excess in the first two bins. The extension of the first bin to $\rm -1000\; km\;s^{-1}$ is justified and motivated by the need to take into account uncertainties in the systemic redshift and to consider possible inflows. We also compute the average occurrence of CIV absorptions in bins of  $\rm 5000 \;km\,s^{-1}$ excluding only the first two velocity bins, obtaining $37\pm5$, and the excess is still significant ($\sim 6\,\sigma$). Most of the excess in the first bin can be explained by the subset of CIV absorptions with detected NV. Indeed, if we exclude those systems, we obtain a number density of 38, that is within 1.32$\,\sigma$ of the average value. The equivalent values corresponding to weak ($\rm W_0  < 0.2$ \AA\,) and strong ($\rm W_0  > 0.2$ \AA\,) lines are reported in Table~\ref{tab_occurrence}.

Hence, the excess is real over the random incidence for the whole sample of CIV NALs.  When divided by equivalent width (see the bottom panel of Fig.~\ref{offset_vel}) CIV absorbers show a significant excess over the random occurrence at $\rm 5000 <v_{abs}< 10,000\; km\,s^{-1}$ for both weak and strong lines. While, at $\rm v_{abs} < \rm 5000\; km\,s^{-1}$ the excess is more significant for the strong lines. We will come back to this point in section $\S5.1$

The excess occurs also in the SiIV velocity offset distribution (see Fig.~\ref{offset_vel}, right panels). As was the case for CIV, the full SiIV sample shows an excess over a random incidence in the first two velocity bins ($< \rm 10,000\; km\,s^{-1}$). 
We would like to highlight the different shape of the two distributions for  the strong lines. The SiIV excess is less significant over the random incidence with respect to the CIV one.
The possibility that these excesses are due to the environment surrounding the host galaxies of luminous quasars is discussed in section $\S 5$. 
All the computed numbers for the SiIV velocity offset distributions are reported in Table~\ref{tab_occurrence}.

\begin{table}
\caption{Occurrence of NALs in the CIV and SiIV velocity offset distributions.}
\begin{minipage}{75mm}
\label{tab_occurrence}
\begin{tabular}{c c  c c }
\hline
 {}&Random$^a$  & Excess$^b$&  Significance$^c$\\
 \hline
 All  CIV sample&$32.2\pm4.4$&$59.5; 64.2$&$6.5\sigma; 7.6\sigma$\\

 $\rm W_{1548} > 0.2$ \AA &$13.2\pm2.9$ &$29.0; 18.6$&$5.5\sigma; 1.9\sigma$\\

$\rm W_{1548} < 0.2$ \AA & $17.5\pm3.3$&$30.5; 45.6$&$4.0\sigma; 8.6\sigma$\\
 \hline

 All SiIV sample&$13.6\pm2.8$&$22.5; 26.4$&$3.1\sigma; 4.5\sigma$\\
$\rm W_{1393} > 0.2$ \AA &$4.6\pm1.7$&$8.5; 5.4$&$2.3\sigma; 0.5\sigma$\\
$\rm W_{1393} < 0.2$ \AA &$9.0\pm2.3$&$14.0; 21.0$&$2.1\sigma; 5.2\sigma$\\
\\
\end{tabular}
$^a$ Average number density of NALs in bin of $\rm 5000\; km\, s^{-1}$ away from the $\rm z_{em}$ ($\rm v_{abs}>5\times 10^4\;km\,s^{-1}$, for CIV and $\rm v_{abs}>2\times 10^4\;km\,s^{-1}$, for SiIV);\\ $^b$ number density of NALs in the first two bins: $\rm -1000 < v_{abs}< 5000\; km\, s^{-1}$ and $\rm 5000 < v_{abs}<10,000 \; km\, s^{-1}$;\\ $^c$ significance of the excess over the random occurrence.
\end{minipage}
\end{table}

Next we investigate the possible dependence of the NAL velocity offset distributions on the quasar bolometric luminosity. The results are presented for CIV and SiIV in Fig.~\ref{offset_vel_L}, left and right plots, respectively. To this aim the sample is split in two according to the quasar bolometric luminosity. In both diagrams the upper panel shows the velocity offset distribution for the whole sample of lines, and the middle and the bottom panels show the results for lines with $\rm W_{0}>0.2$ \AA\,and $\rm W_{0}<0.2$ \AA, respectively. Color-shaded histograms represent the lines detected in quasars with luminosity $\rm log\,(L_{bol})<46.8$ $\rm erg\,s^{-1}$, while the black hashed histogram shows those in quasars with $\rm log\,(L_{bol})>46.8$ $\rm erg\,s^{-1}$.

\begin{figure}
   \centering
   \includegraphics[width=\hsize]{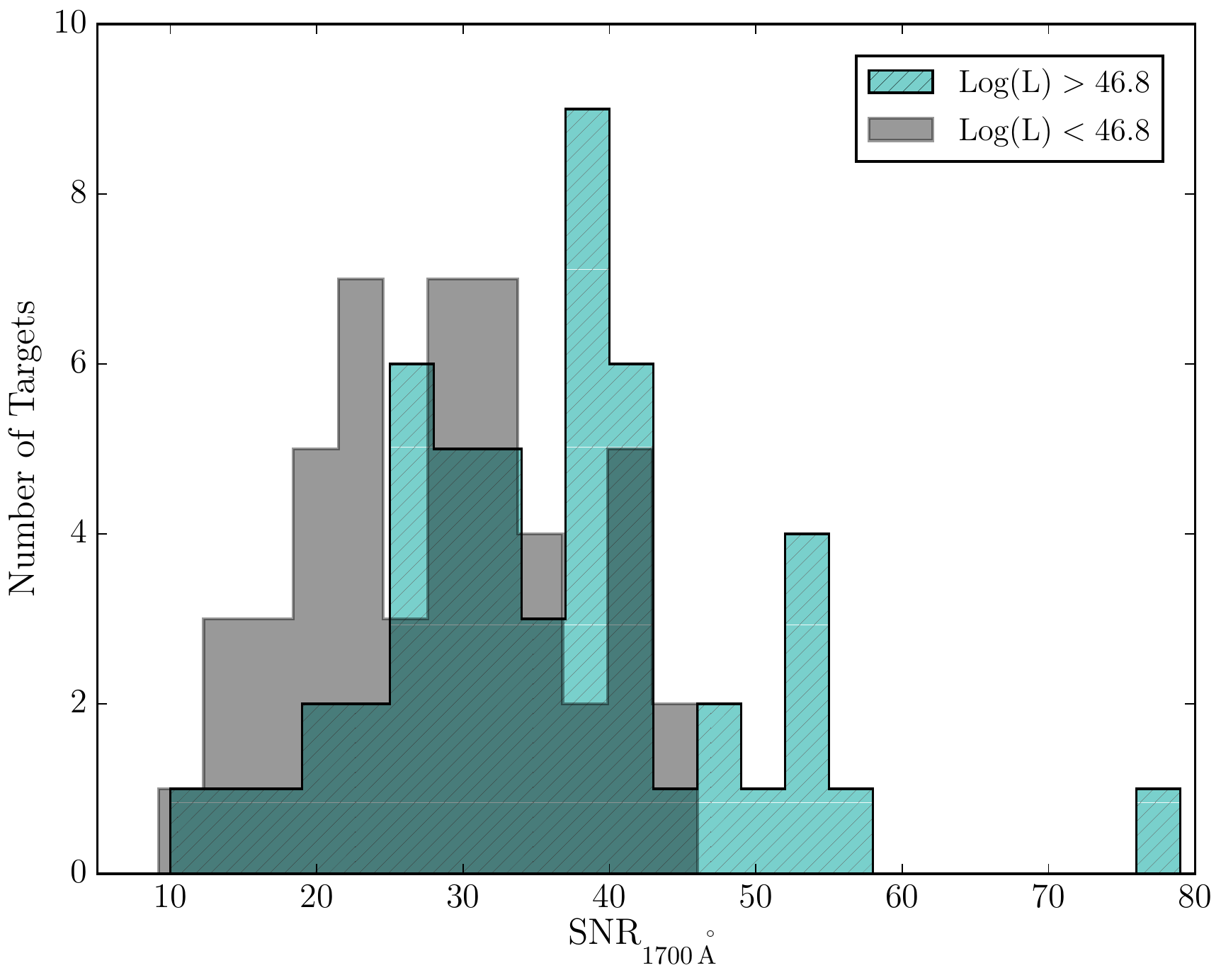}
     \caption{S/N (average pixel S/N near rest-frame wavelength 1700 \AA) distribution of XQ-100 targets. The shaded histogram represents the quasars with $\rm log(L_{bol})>46.8$, the diagonal filled histogram represents those with $\rm log(L_{bol})<46.8$.
             }
             \label{snr_test}
  \end{figure}

 In both of the upper panels, the more luminous objects exhibit a slightly larger fraction of absorbers. We explore the possibility that this trend is due to the different S/N between the quasar spectra. Although the S/N in our sample is quite homogeneous, the most luminous quasars have slightly larger S/N spectra, as shown in Fig~\ref{snr_test}. If the observed effect is due to S/N, we should not see the same behavior of the velocity offset distributions for the stronger lines, that are less affected by the different S/N. Indeed, we do not see any difference related with the bolometric luminosity for those lines (see middle panels of Fig.~\ref{offset_vel_L}). The weaker lines show the same slight dependence from the bolometric luminosity we found in the whole sample (see lower panels of Fig.~\ref{offset_vel_L}).  
Thus, there is no evidence of a correlation between the incidence of absorbers observed in the offset velocity distribution and the quasar bolometric luminosity for both CIV and SiIV. The difference in Fig.~\ref{offset_vel_L} can be ascribed to the slightly higher S/N of the spectra of the brighter objects. This effect with S/N was seen before for MgII absorbers (e.g., Lawther et al. 2012; M{\'e}nard et al. 2008).

To  summarize, we detect a $\rm \sim 8\,\sigma$ excess of NALs within $\rm 10,000\;km\,s^{-1}$ of the $\rm z_{em}$ over the random occurrence of NALs. This excess is expected (Foltz et al. 1986; Ganguly et al. 2001; Vestergaard 2003), but extends to somewhat larger velocities than the standard $\rm 5000 \; km\, s^{-1}$ velocity cut-off adopted to define associated system, in general agreement with the results of Weymann et al. (1979), Misawa et al. (2007), Tripp et al. (2008), Wild et al. (2008) and Nestor et al. (2008). This common used cut-off is determined for high value of $\rm W_{1548}$ NALs and if we limit our analysis to $\rm W_{1548}>0.2$ \AA\, our results are consistent with those of previous works. The extended velocity range of associated NALs is seen especially for the weaker lines $\rm W_{1548}<0.2$\,\AA\, (but above our detection limit, 0.015\,\AA).
Based on this excess frequency, we adopt $\rm 10,000\;km\,s^{-1}$ as the nominal boundary between intervening and associated systems. Because this velocity range of associated absorbers is statistically motivated, additional associated NALs may be present at $\rm v_{abs}>10,000 \;km\,s^{-1}$, just as there may be some intervening NALs in the excess of lines with respect the average value at $\rm v_{abs}<10,000 \;km\,s^{-1}$.

\subsection{Absorber number density evolution $\rm dn/dz$}

As a final test we want to assess whether the excess we find is due to the NAL redshift evolution. We estimate the number of absorbers in the considered range of velocity offset that are expected due to intervening structures at the same redshift.
We calculate the number density evolution $\rm dn/dz$ splitting our sample into two: the associated CIV lines with $\rm v_{abs}<10,000\; km\;s^{-1}$ and the intervening ones with $\rm v_{abs}>10,000\; km\;s^{-1}$.

\begin{figure}
  \centering
  \includegraphics[width=8cm]{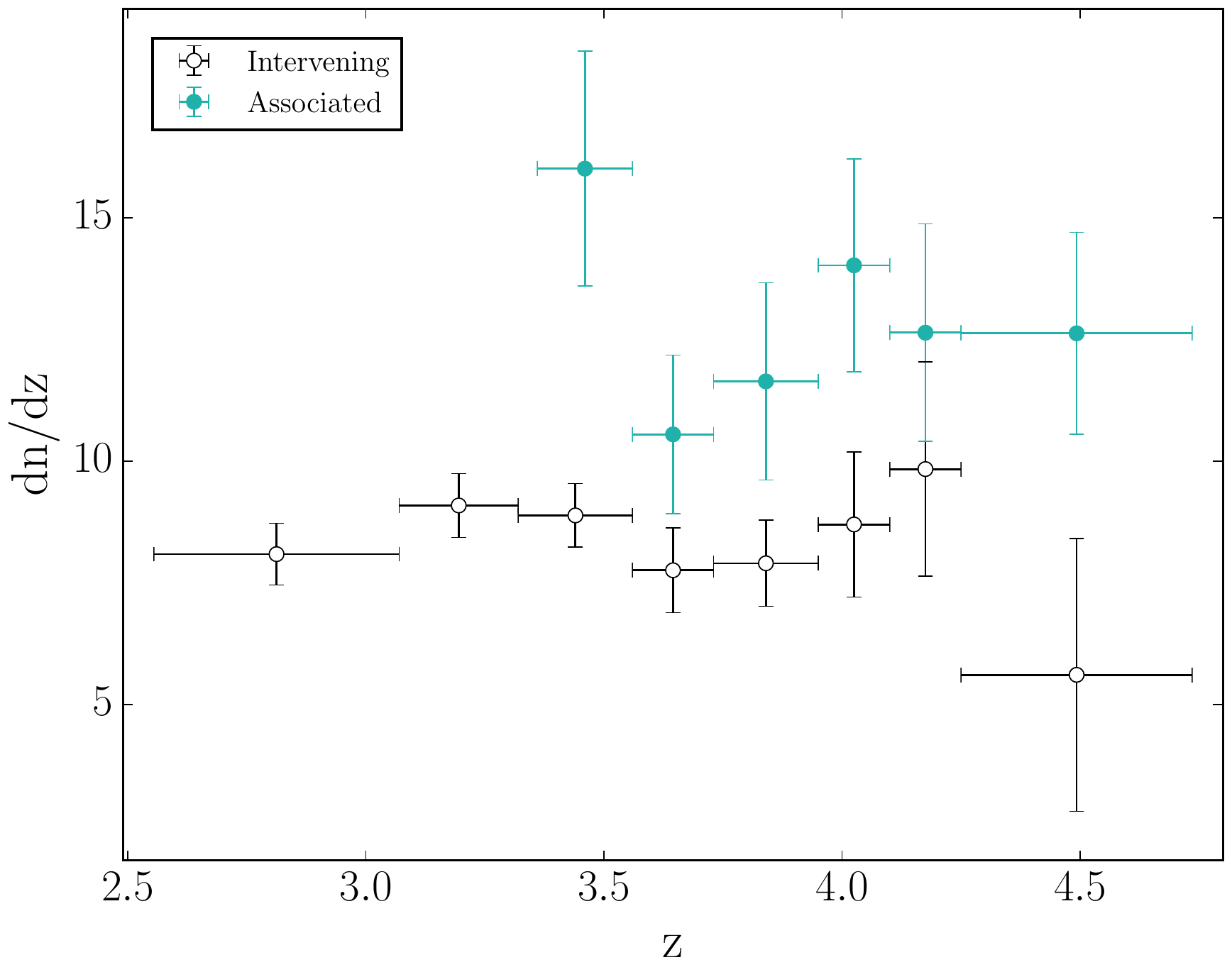}
  \caption{Absorber number density evolution dn/dz of the whole sample of CIV systems. Filled circles represent associated systems with $\rm v_{abs}<10,000\; km\;s^{-1}$ while open circles  the intervening ones with $\rm v_{abs}>10,000\; km\;s^{-1}$. The error bars represent the propagation of the Poissonian uncertainties.}
    \label{dn_dz}
\end{figure}

The absorber number $\rm n(z)$ is measured by counting the CIV absorption lines for a given equivalent width range in the considered redshift range for each line of sight. The line count $\rm n$ is then divided by the covered redshift interval $\rm \Delta z$ to obtain $\rm dn/dz$.
The result is presented in Fig.~\ref{dn_dz}. 
We can see that the excess of the associated lines over the intervening ones is present in all covered redshift bins although its significance varies from bin to bin. This result supports the fact that the observed excess in the offset velocity distribution is not an effect of the NAL redshift evolution, but is due to quasar environment. This test is done only for CIV because contamination by the $\rm Ly\alpha$ forest prevented us from investigating the same wavelength coverage for other ions.

 \subsection{Covering Fraction of the studied ions}

\begin{table*}
\caption{Covering fractions of CIV, SiIV, NV and CII.}             
\centering          
\begin{tabular}{c c c c c}   
\hline\hline       
Velocity [$\rm km\,s^{-1}$] &$\rm f_C^{1548}$ &$\rm f_C^{1393}$ &$\rm f_C^{1238}$ &$\rm f_C^{1334}$\\
\hline
&&$\rm W_{0}>0.2 $\AA &&\\
\hline                    
  $\scriptstyle\rm v_{abs}<5000$              & $0.43\pm0.07$ & $0.13\pm0.04$ & $0.22\pm0.05$ & $0.08\pm0.03$ \\      
  $\scriptstyle\rm 5000<v_{abs}<10,000$& $0.27\pm0.05$ & $0.09\pm0.03$ & -                         & $0.02\pm0.01$ \\
   $\scriptstyle\rm v_{abs}<10,000$           & $0.43\pm0.07$ & $0.13\pm0.04$ & -                        & $0.10\pm0.3$ \\
\hline
&&Whole sample&&\\
\hline                    
  $\scriptstyle\rm v_{abs}<5000$              & $0.72\pm0.09$ & $0.32\pm0.04$ & $0.33\pm0.06$ & $0.15\pm0.04$ \\    
  $\scriptstyle\rm 5000<v_{abs}<10,000$& $0.64\pm0.08$ & $0.39\pm0.03$ & -                         & $0.09\pm0.03$ \\
   $\scriptstyle\rm v_{abs}<10,000$           & $0.72\pm0.08$ & $0.32\pm0.04$& -                         & $0.22\pm0.05$ \\
\hline   
               
\end{tabular}

\label{cov_fra}
\end{table*}

  \begin{figure*}
  \centering
  \includegraphics[width=8cm]{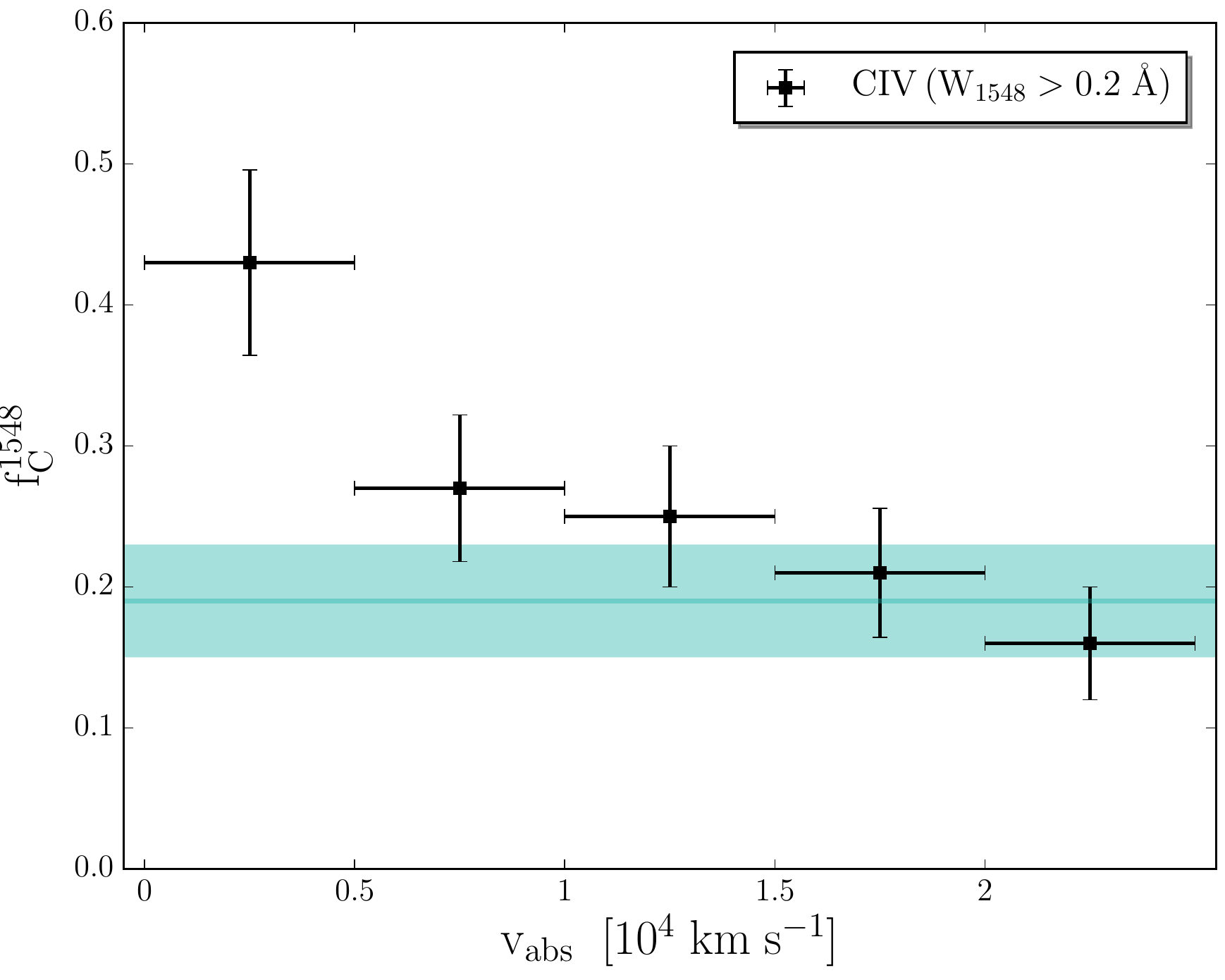}
  \qquad\qquad
  \includegraphics[width=8cm]{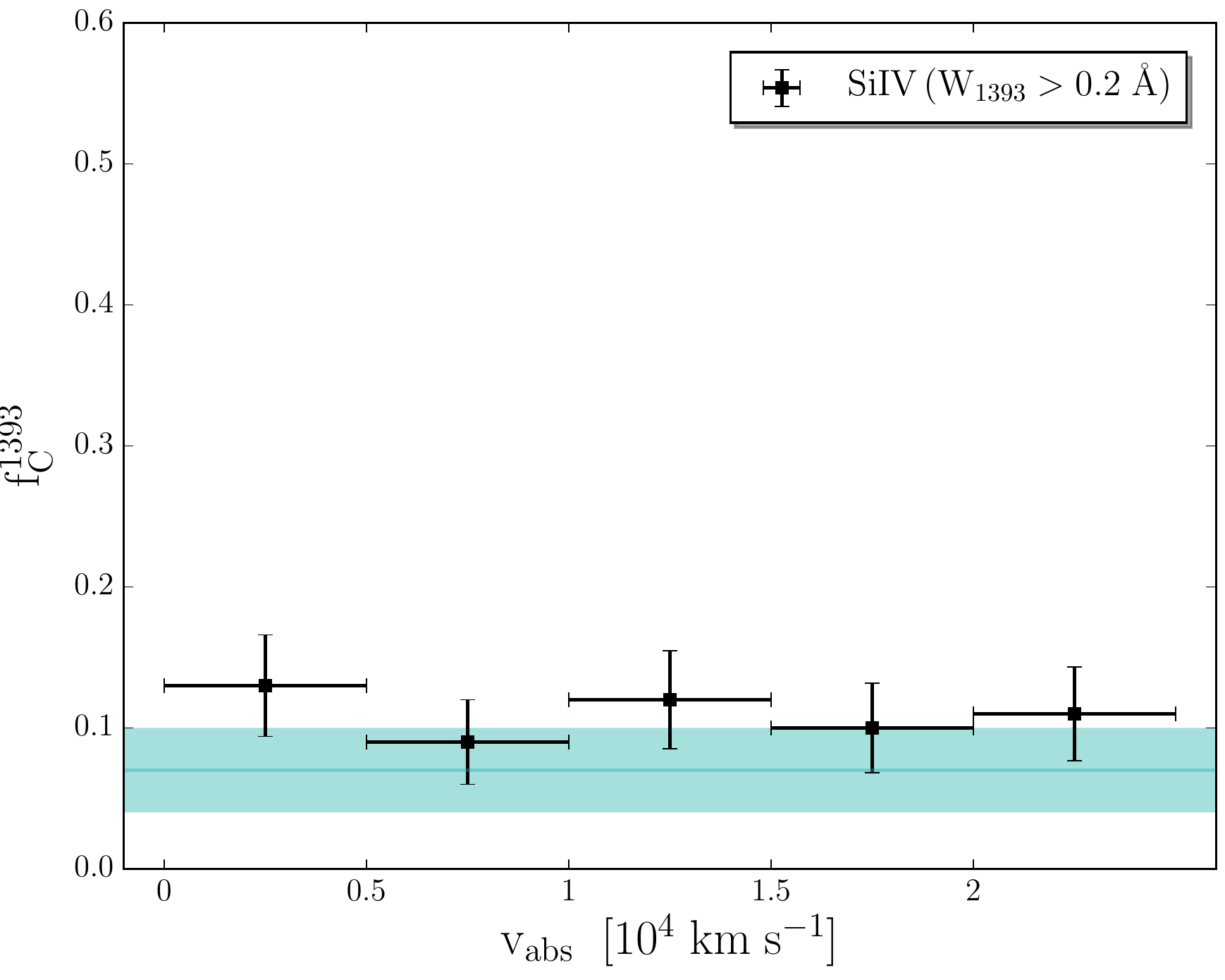}
  \caption{Covering fractions, $\rm f_C$, for $\rm CIV_{1548}$ (left) and $\rm SiIV_{1393}$ (right) estimated from the fraction of quasars exhibiting at least one absorber with $\rm W_0>0.2$ \AA, in bins of velocity offset from $\rm z_{em}$ (horizontal error bars). The horizontal solid line in both plots shows the mean covering fraction for $\rm W_0>0.2$ \AA\,lines measured for random $\rm 5000 \; km\, s^{-1}$ intervals far from the quasars, the shaded band represents the $\rm 1\, \sigma$ error.}
   \label{first_cov}
\end{figure*}

  \begin{figure*}
  \centering
  \includegraphics[width=8cm]{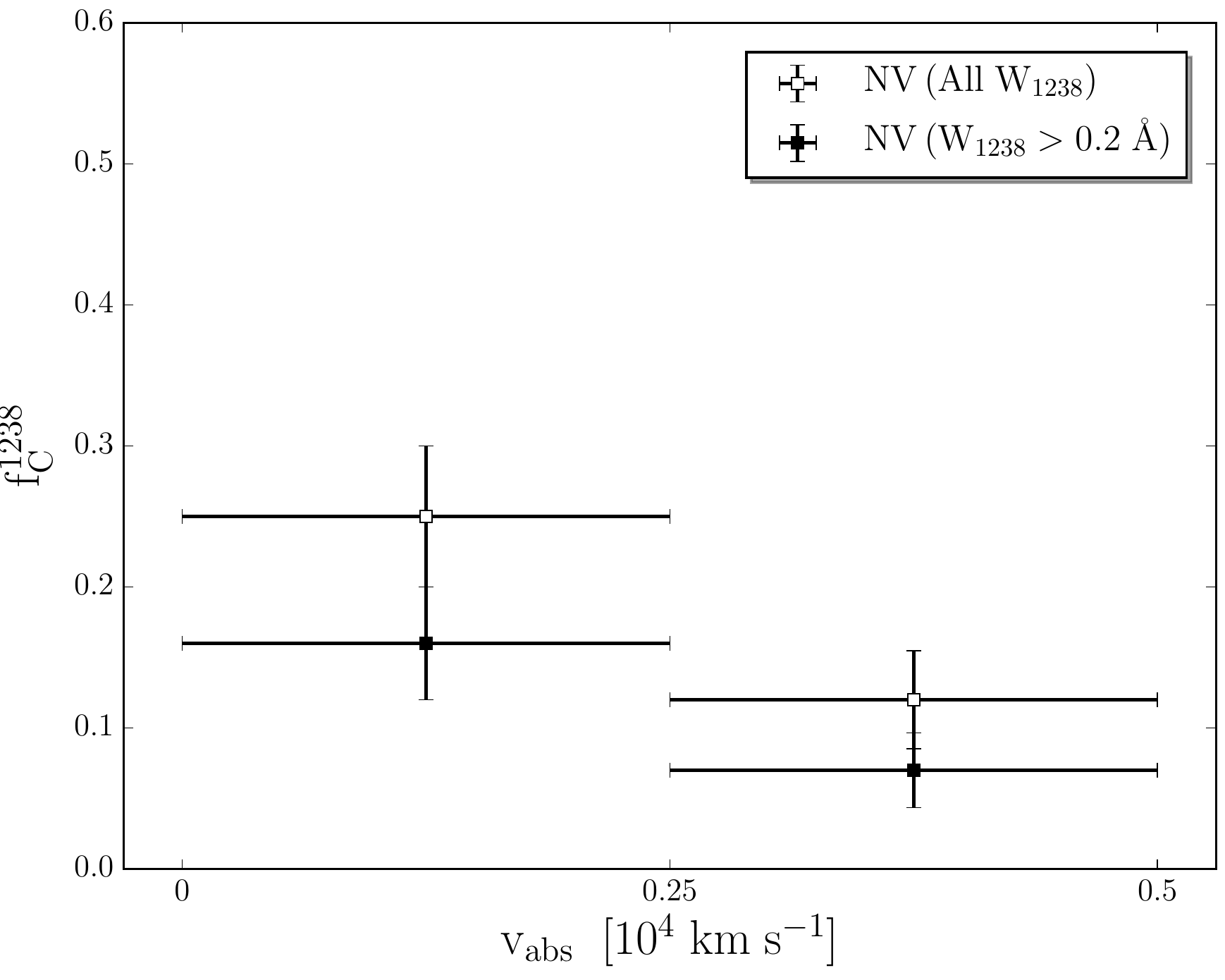}
  \qquad\qquad
  \includegraphics[width=8cm]{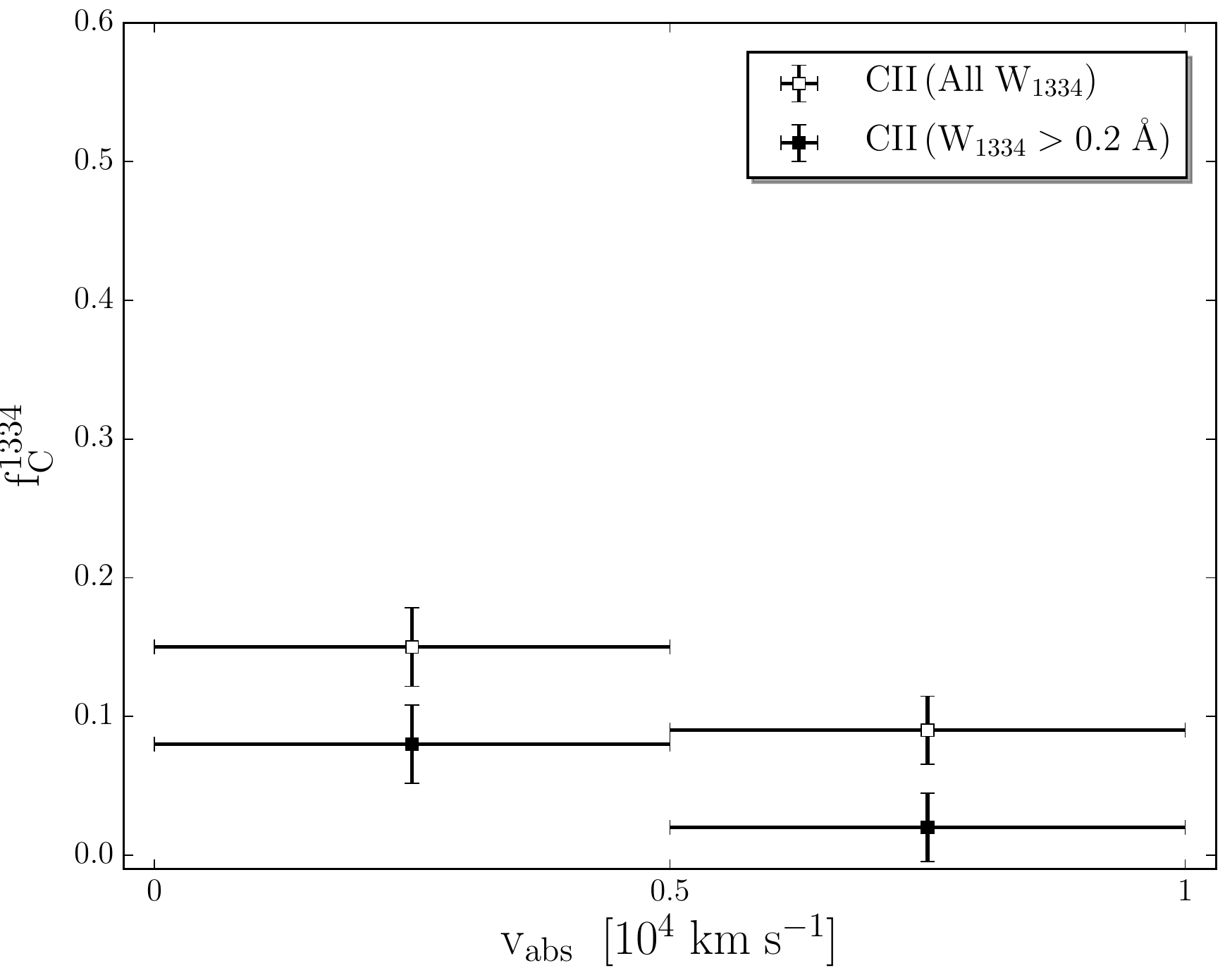}
  \caption{Covering fractions, $\rm f_C$, for $\rm NV_{1238}$ (left) and $\rm CII_{1334}$ (right) estimated from the fraction of quasars exhibiting at least one absorber with $\rm W_0>0.2$ \AA\,(filled circles) or for the whole sample (open circles), in bins of velocity offset from $\rm z_{em}$.} 
   \label{second_cov}
\end{figure*}

To better characterize the environment close to our quasar sample, we measure the covering fractions, $\rm f_C$, of the studied ions as a function of the velocity offset from $\rm z_{em}$. These are defined as the ratio between the number of quasars exhibiting at least one absorber with $\rm W_0>0.2$ \AA\, within a given velocity separation of $\rm z_{em}$ and the total number of quasars. Poissonian uncertainties are taken into account. The complete list of covering fraction measurements is reported in Tab.~\ref{cov_fra}. We decide to consider lines with $\rm W_0>0.2$ \AA\,to better compare our results with previous works present in the literature (see section $\S5$), but measurements of the $\rm f_C$ based on the whole sample are also presented in Tab.~\ref{cov_fra}.

Results are shown in Fig.~\ref{first_cov} for $\rm CIV_{1548}$ (left) and $\rm SiIV_{1393}$ (right), compared with the random occurrence computed as the fraction of quasars showing at least one absorber with $\rm W_0>0.2$ \AA\,at large velocity separation from $\rm z_{em}$ ($\rm v_{abs} > 5 \times 10^4\, km \,s^{-1}$, for CIV and $\rm v_{abs} > 2 \times 10^4\, km \,s^{-1}$, for SiIV). 
The CIV covering fraction, $\rm f_C^{1548}$, remains statistically significant over the random occurrence beyond $\rm v_{abs}\approx10,000\; km\, s^{-1}$. Then, it shows a shallow declining incidence. This suggests that most of the excess of CIV gas at low velocities lies within the host halo. Interestingly, $\rm f_C^{1548}$ for $\rm v_{abs}<5000 \; km\, s^{-1}$ and for $\rm v_{abs} <10,000 \; km\, s^{-1}$ have the same value. Furthermore, we confirm that quasars showing an absorber with $\rm 5000<v_{abs}< 10,000\; km\, s^{-1}$ always have an absorber with $\rm v_{abs}<5000 \; km\, s^{-1}$, but the opposite is not always true. This points out the presence of a complex velocity structure, increasing the probability of these absorbers to be part of outflows. We note also that considering the whole sample (Tab.~\ref{cov_fra}) almost 3/4 ($\sim 72$ percent) of the quasars show CIV NALs within $\rm 10,000\; km\, s^{-1}$. This is comparable to the frequency ($\gtrsim 60$ percent) observed by Vestergaard (2003) and shows how common this phenomenon is. Vestergaard et al. (2003) also found that about 25 percent of the quasars have associated NALs with $\rm W_{1548}>0.5$ \AA.   
 To allow a more direct and quantitative comparison with this work, we can consider in our sample only the absorbers with $\rm W_{1548}>0.5$ \AA\, and within $\rm v_{abs}<5000\; km\;s^{-1}$. Indeed we do find exactly 25 percent.
For SiIV, $\rm f_C^{1393}$ is marginally statistical significant over the random distribution only at $\rm v_{abs}<5000\; km\, s^{-1}$. In fact, we have already seen in Fig.~\ref{offset_vel} that the SiIV excess is largely dominated by weak lines, with $\rm W_{1393}<0.2$ \AA.

The covering fractions measured for  $\rm NV_{1238}$ (left) and $\rm CII_{1334}$ (right) are presented in Fig.~\ref{second_cov}. 
The fraction $\rm f_C^{1238}$ (whole sample) is fairly high $\rm \sim 0.25$ at $\rm v_{abs}<2500\; km\, s^{-1}$, then it drops steeply to $\rm \sim0.12$ at $\rm 2500<v_{abs}< 5000\; km\, s^{-1}$. Since the velocity range in which we have been able to explore the presence of NV is relatively narrow because of the contamination by the $\rm Ly\alpha$ forest, we do not have a reference value to compare the  $\rm f_C^{1238}$ with. 

Fechner \& Richter (2009, FR09, hereafter), have carried out a survey of NV systems with a sample of 19 higher resolution spectra of quasars with $\rm 1.5\lesssim z\lesssim2.5$. They find that the fraction of intervening CIV systems showing NV absorption is $\rm \sim 11$ percent, without any cut in equivalent width, while roughly 37 percent of associated CIV absorbers ($\rm v_{abs}<5000\,km\,s^{-1}$) exhibit NV as well. 
They do not say how many quasars show at least one intervening NV thus, we could take their $\rm \sim 11$ percent as an upper limit of the random occurrence and our detection frequency  would remain statistically significant. The steep drop of the $\rm f_C^{1238}$at $\rm v_{abs}>2500\; km\, s^{-1}$ close to values that are typical of intervening regions suggests that the NV gas with $\rm v_{abs}<2500\; km\, s^{-1}$ lies predominantly inside the host halo. 
In addition, we can directly compare to their 11 percent the fraction of CIV, with $\rm v_{abs}<5000\; km\, s^{-1}$, exhibiting NV in our sample, that is $\rm \sim 35$ percent. 
This is strong evidence that NV systems exhibit an excess above random occurrence.
However, we have to take into account that the sample analyzed by FR09 is most likely incomplete due to blending with the $\rm Ly\alpha$ forest. So, the actual rate of incidence might be higher, in particular for low column density features. We will discuss in section $\S 5$ the possibility of different origins for intervening and associated NV systems, and the improvements that could be made to verify this hypothesis. 

The CII covering fraction, $\rm f_C^{1334}$, is smaller than that of NV. If we consider only the bin with $\rm v_{abs}<5000\; km\, s^{-1}$, $\rm f_C^{1238}=0.33$ while $\rm f_C^{1334}=0.15$. Interestingly, we notice that  the CII detections with  $\rm W_{1334}>0.2$ \AA\,are all corresponding to damped Ly-alpha (DLA and sub-DLA) systems (Berg et al. 2016; submitted). DLAs are defined to be systems where the hydrogen column density is larger than $\rm N_{HI}>2\times10^{20}\;cm^{-2}$. We would like to emphasize that CII detections in our sample are probably not tracing gas sited in the host halo.

\subsection{Investigating the ionization structure of the absorbers}

  \begin{figure*}
  \centering
  \includegraphics[width=8cm]{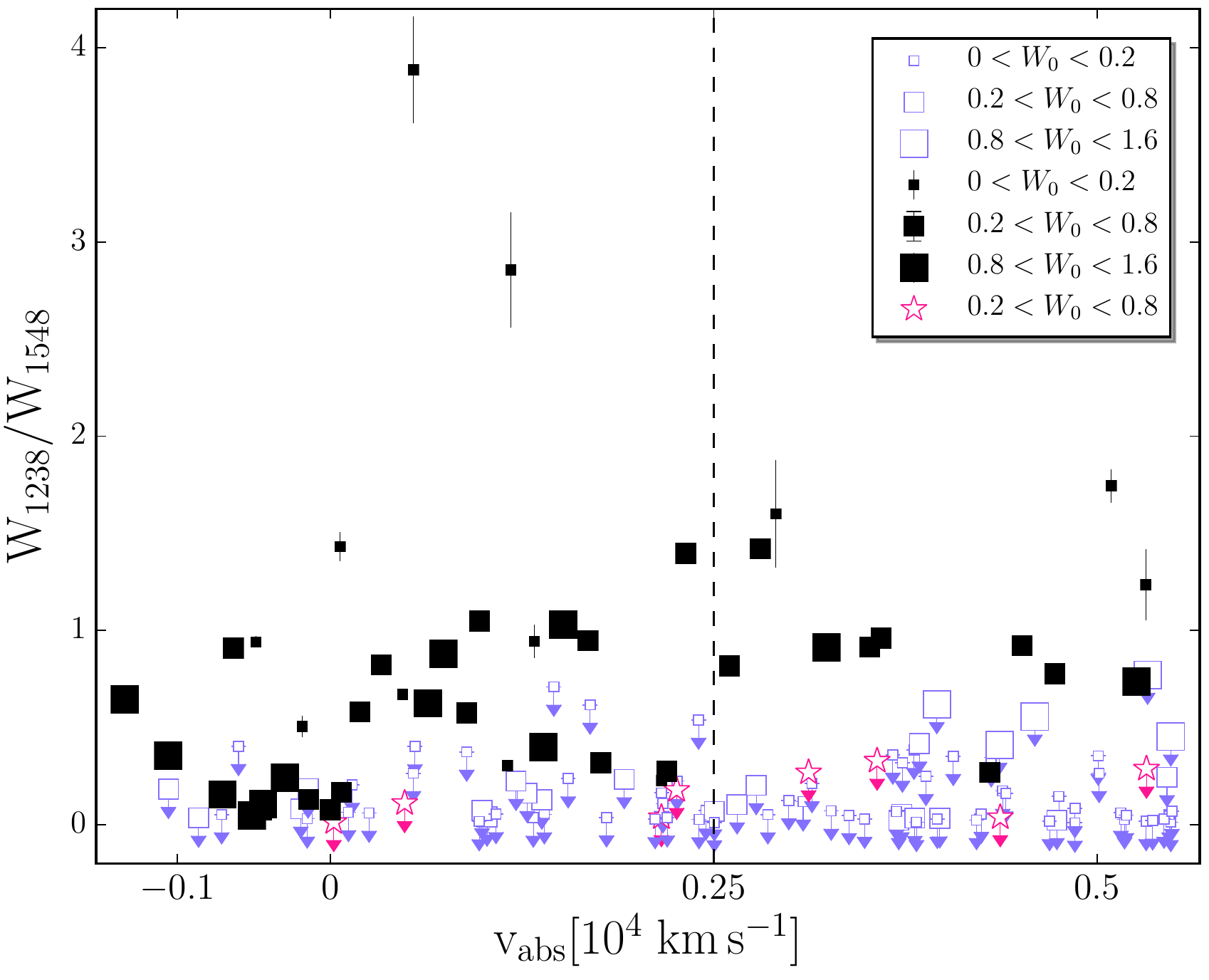}
  \qquad\qquad
  \includegraphics[width=8cm]{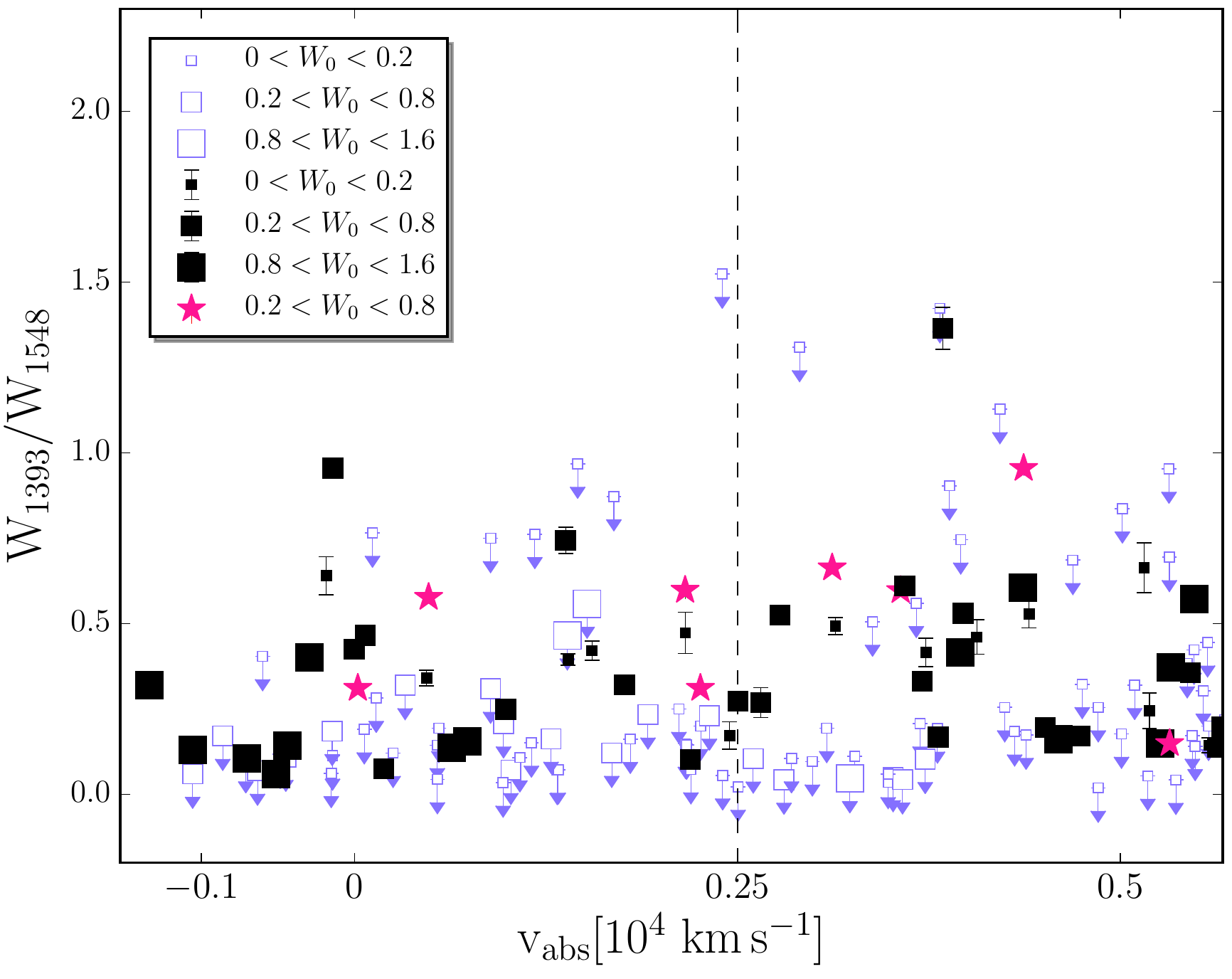}
  \caption{Velocity offset distribution with respect to the quasar $\rm z_{em}$ of the NV and CIV (left), SiIV and CIV (right) equivalent width ratio. The size of the symbol represents the corresponding $\rm W_{1548}$ \AA\, strength, indicated by the legends. Detections ($\rm 3\sigma$) are the filled squares. The open squares show the rest-frame equivalent width $\rm 3\sigma$ upper limit values for the non-detections. The damped Ly-alpha (DLA) systems present in our sample are marked by open stars (filled stars) in the left panel (right panel) and they show NV upper limits  (SiIV detections). The vertical dashed lines represent the separation of the velocity range in two bins.}
      \label{ratio_ions}
\end{figure*}

  \begin{figure}
  \centering
  \includegraphics[width=8cm]{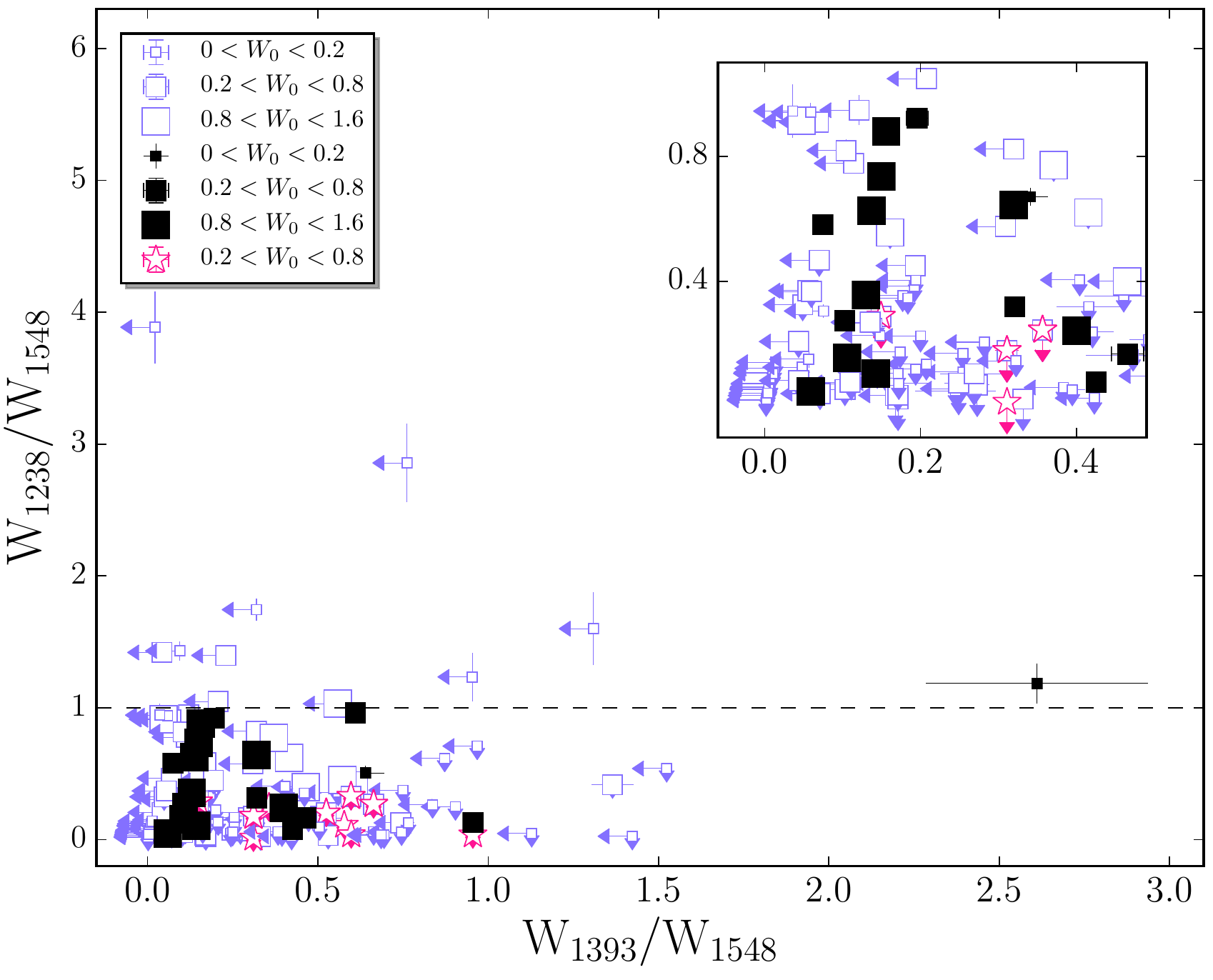}
  \caption{Equivalent width ratio between NV and CIV versus the one of SiIV and CIV. Same meaning of the symbols as in Fig.~\ref{ratio_ions}. The inset represents a zoom of the bottom left region of the diagram. The horizontal dashed line represents the equality of the NV and CIV equivalent widths.}
    \label{match}
\end{figure}

Historically, it has been very difficult to separate intrinsic NALs from intervening ones without high-resolution spectra because of the similarities of their line profiles. Since one of our goals is to determine additional criteria that can be used to disentangle the two classes of absorbers, it is interesting to consider the relative number of NALs in different transitions and their velocity offset distributions. The ion that better traces the effects of the quasar ionization field will offer the best statistical means of identifying intrinsic systems.

The velocity offset distribution with respect to the quasar $\rm z_{em}$ of the ratios of NV and CIV (SiIV and CIV) equivalent width is shown in Fig.~\ref{ratio_ions}, left (right) panel. As already described in $\S 3.2$, we have inspected all the spectra of our sample looking for NV and SiIV absorptions at the same CIV redshifts. NV is only measured out to $\rm 5500\; km\,s^{-1}$ from $\rm z_{em}$, hence the smaller velocity range of the diagram relative to previous ones.
We search for a correlation between the offset velocity and the equivalent width of CIV and NV NALs. 
It is clear from Fig.~\ref{ratio_ions} (left) that CIV is equally distributed within the velocity range in which it has been possible to look for NV absorptions, even if the strongest systems preferentially reside at zero velocity shift. We divide the velocity range into two bins, the first one with $\rm -1000\leqslant v_{abs}\leqslant2500\; km\;s^{-1}$ and the second one with $\rm 2500 \leqslant v_{abs}\leqslant 5500 \; km\;s^{-1}$. We find 72 CIV systems in the first bin and 68 in the second one. Most of the NV detections are located closer to the emission redshift: we identify 34 NV detections and 38 upper limits in the first bin ($\rm -1000 <v_{abs}< 2500\; km\,s^{-1}$) and 12 detections and 56 upper limits within the second bin ($\rm 2500 <v_{abs}< 5500\; km\,s^{-1}$). In other words, looking at the CIV absorbers in the first bin we have a probability of 47 percent to find a NV at the same absorption redshift, while this probability decreases to 18 percent in the second bin. This trend can be explained if we consider the effect of the quasar ionizing radiation. Indeed, closer to the emission redshift the radiation field is more intense and it allows nitrogen to be highly ionized. 
As expected DLA systems (marked by open stars in the figure) do not have any detected associated NV absorption systems because of the self-shielding effects of $\rm HI$, that prevents highly ionized atoms to be formed (but see Fox et al. 2009).

We search also for a correlation between the offset velocity and the equivalent width ratio of CIV to SiIV NALs in the same velocity range. In Fig.~\ref{ratio_ions} (right), SiIV detections do not preferentially reside around $\rm z_{em}$, nor do we find a clustering of the strongest systems close to $\rm z_{em}$. We find 26 detections and 46 upper limits in the first bin ($\rm -1000 <v_{abs}< 2500\; km\,s^{-1}$); and  27 and 42, respectively,  in the second one ($\rm 2500 <v_{abs}< 5500\; km\,s^{-1}$). This corresponds to a probability of 36 percent to find a SiIV at the same redshift as CIV absorption in the first bin and a probability of 40 percent in the second one. Above unity, representing the equality of the equivalent widths ratio, we find only 2 SiIV detections. Indeed, there are not many systems with very strong SiIV absorption able to dominate over the CIV absorption. The DLA systems, indicated by the filled stars in the figure, exhibit SiIV detections and their values are distributed over the entire range of SiIV equivalent widths.

Bearing in mind the different behavior of NV and SiIV ionization paths, we match the results of the latter two figures. The result is shown in Fig.~\ref{match}. Immediately, we can see that above the dashed horizontal line, representing the equality of the NV and CIV equivalent widths, there is only one system with a SiIV detection. When a strong NV is present, we find mainly upper limits for the SiIV. Conversely, the DLA systems show detections for the SiIV and only upper limits for the NV. We do not find a large number of systems showing both ions and most of them have SiIV with $\rm W_{1393}<0.2$ \AA\,(see Fig.~\ref{match}). Interestingly many of them (9 out of 19) are at the same redshift as a strong CIV NAL, with $\rm W_{1548}>0.8$ \AA. Once more, we can appreciate the effects of the quasar ionizing radiation. Since the NV and SiIV have very different ionization potentials, 97.9 and 33.5 eV respectively, it is easier to find NV absorbers with respect to SiIV where the radiation field is stronger and vice versa. Therefore, this result confirms that NV is a better tracer of the hard quasar spectra.

\section{Discussion}

Quasar outflows have been increasingly invoked from theoretical models of galaxy formation and evolution to regulate both the star formation in the host galaxies and the infall of matter towards the central SMBH (Granato et al. 2004; Di Matteo, Springel \& Hernquist 2005; Hopkins \& Elvis 2010). 
The SMBH at the center of galaxies can produce a terrific amount of energy ($\rm \sim 10^{62}\;erg$). Even if just a few percent of the quasar bolometric luminosity were released into the ISM of the host galaxy, it would represent significant feedback for the host galaxy evolution, (e.g. $\rm \sim 5$ percent $L_{bol}$, Scannapieco \& Oh 2004; Di Matteo et al. 2005; Prochaska \& Hennawi 2009).
This type of coupling via feedback could provide a natural explanation for the observed mass correlation between SMBHs and their host galaxy spheroids (e.g., King 2003; McConnell et al. 2013).

A key piece of information to understand if outflow feedback can affect the host galaxy evolution, is the fraction of quasar driving outflows, as well as their energetics. The latter quantity can be inferred from the velocity, column density, and global covering factor of the outflowing gas. All these observable quantities are connected with the dynamical models of accretion disk winds. 
 With the aim of studying quasar winds, we must carefully select absorption-line systems that truly sample outflowing gas.

In the systematic search for highly ionized absorption species commonly used to identify outflows (e.g. CIV, SiIV, NV, OVI), the NV ion is detected in previous works with the lowest frequency. However, the fraction of these systems that are intrinsic is much larger than that of other excited-state species. 
Misawa et al. (2007) derived for the NV systems an intrinsic fraction of 75 percent, through partial coverage and line locking methods.
Ganguly et al. (2013), using the same techniques, found an intrinsic fraction of  29-56 percent, and suggested to use NV in constructing large catalogs of intrinsic systems with lower resolution and/or lower S/N data. 
This is not surprising given the overabundance of nitrogen in AGN reported by some authors (e.g., Hamann \& Ferland 1999). Quasars are generally metal-rich (e.g., Dietrich et al. 2003) and about solar $\rm [N/H]$ has been measured in associated systems (e.g., D'Odorico et al. 2004). 
There is also evidence for higher than solar metal abundances in AGN outflows from the analysis of NALs (Wu et al. 2010; Hamann et al. 2011), making NV easily detectable in proximate absorbers. These metallicities are very high and very rare for any intervening absorption system.

Very strong NV absorption is characteristic of many intrinsic NALs, that are known to be related with the AGN winds/host-galaxy because of a velocity within $\rm \sim 5000\;km\,s^{-1}$ of the $\rm z_{em}$, and evidence of partial coverage of the quasar continuum/broad emission line source (Hamann et al. 2000; Srianand et al. 2002, Wu et al. 2010). For example, Wu et al. (2010) addressed the origin of three 2.6<z<3.0 NV absorbers and pointed out how NV is a good estimator of the intrinsic nature of systems. They performed photoionization models to infer the physical conditions for these absorbers, and found metallicities greater than 10 times the solar value, and high ionization parameters ($\rm log\,U\sim0$). The unusual strength of these NV lines resulted from a combination of partial coverage, a high ionization state, and high metallicity.

On this basis, we compute the fraction of quasars hosting at least one intrinsic absorption system exploiting the detection of NV corresponding to CIV absorptions in our sample, to be 33 percent. This value can be compared with the fraction of quasar showing at least one intrinsic NV absorber from previous studies (e.g., $\rm \sim 19$ percent, Misawa et al. 2007). We expect our estimate to be slightly higher than that derived through partial coverage or line-locking, because these selection criteria catch only $\rm 12-44$ percent of intrinsic systems (Ganguly et al. 2013). Indeed, intrinsic absorbers could be too far from the central engine to cause partial coverage, and may not be line locked.

We have shown in the previous section that NV is the ion that traces best (amongst those considered) the effects of the quasar ionization field, offering the best statistical tool to identify intrinsic systems. 
Furthermore, other studies present in the literature have argued that associated NV shows distinctive signatures of its intrinsic nature with respect to intervening systems. 
In particular, FR09 have constructed detailed photoionization models to study the physical conditions of the absorbers and to constrain 
the ionizing radiation field. This work has pointed out that intervening NV absorbers are not tracers of the spectral hardness of the ionizing radiation. These systems were found to be systematically weaker than both CIV and associated NV lines.

 \begin{figure}
  \centering
  \includegraphics[width=8cm]{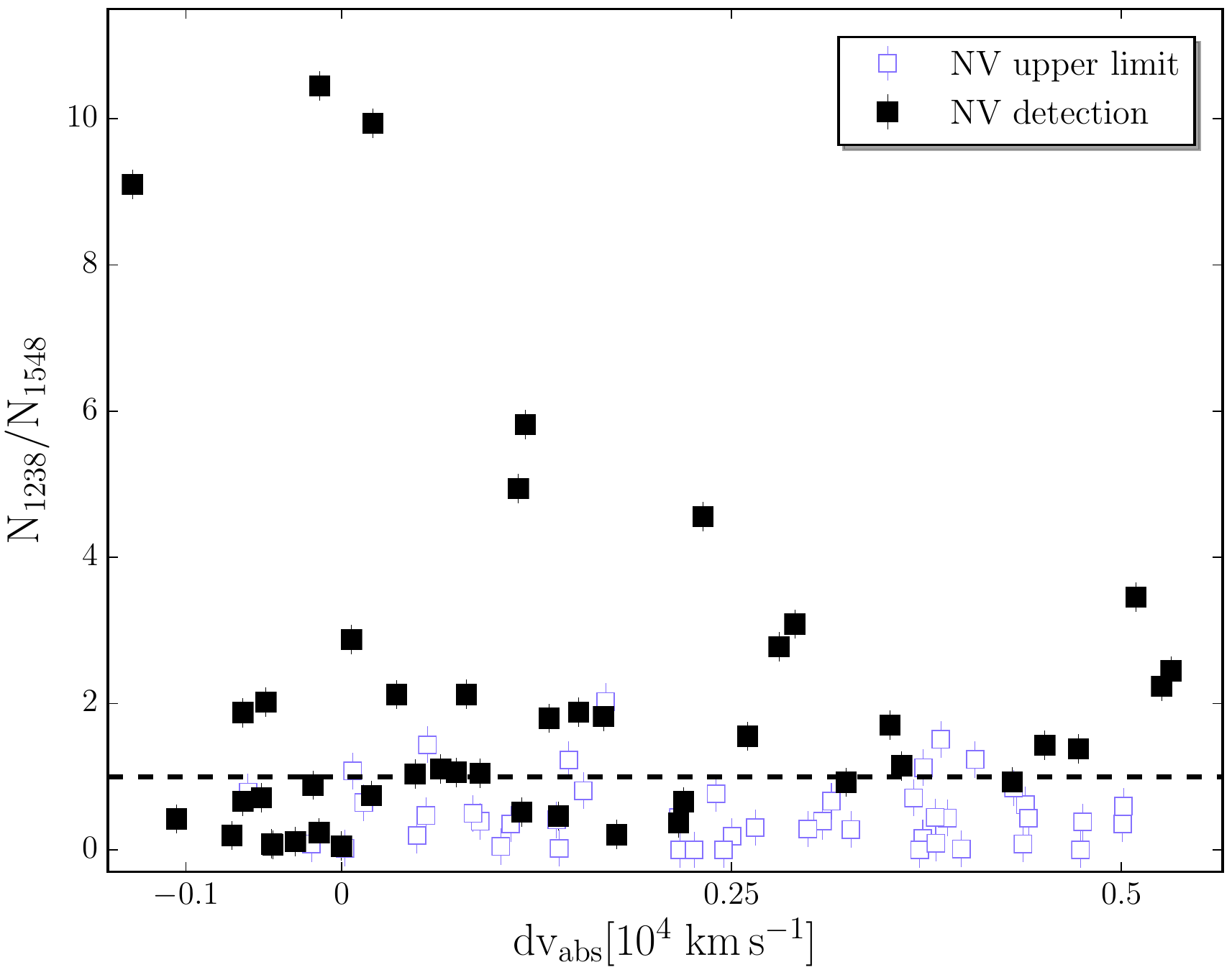}
  \caption{Velocity offset distribution of the NV vs CIV column density ratio. Detections ($\rm 3\sigma$) are marked by the filled squares. The open squares show the column density $\rm 3\sigma$ upper limit values for the non-detections. The dashed horizontal line represents the equality of the ratio.}
    \label{col_ratio}
\end{figure}

  \begin{figure}
  \centering
  \includegraphics[width=8cm]{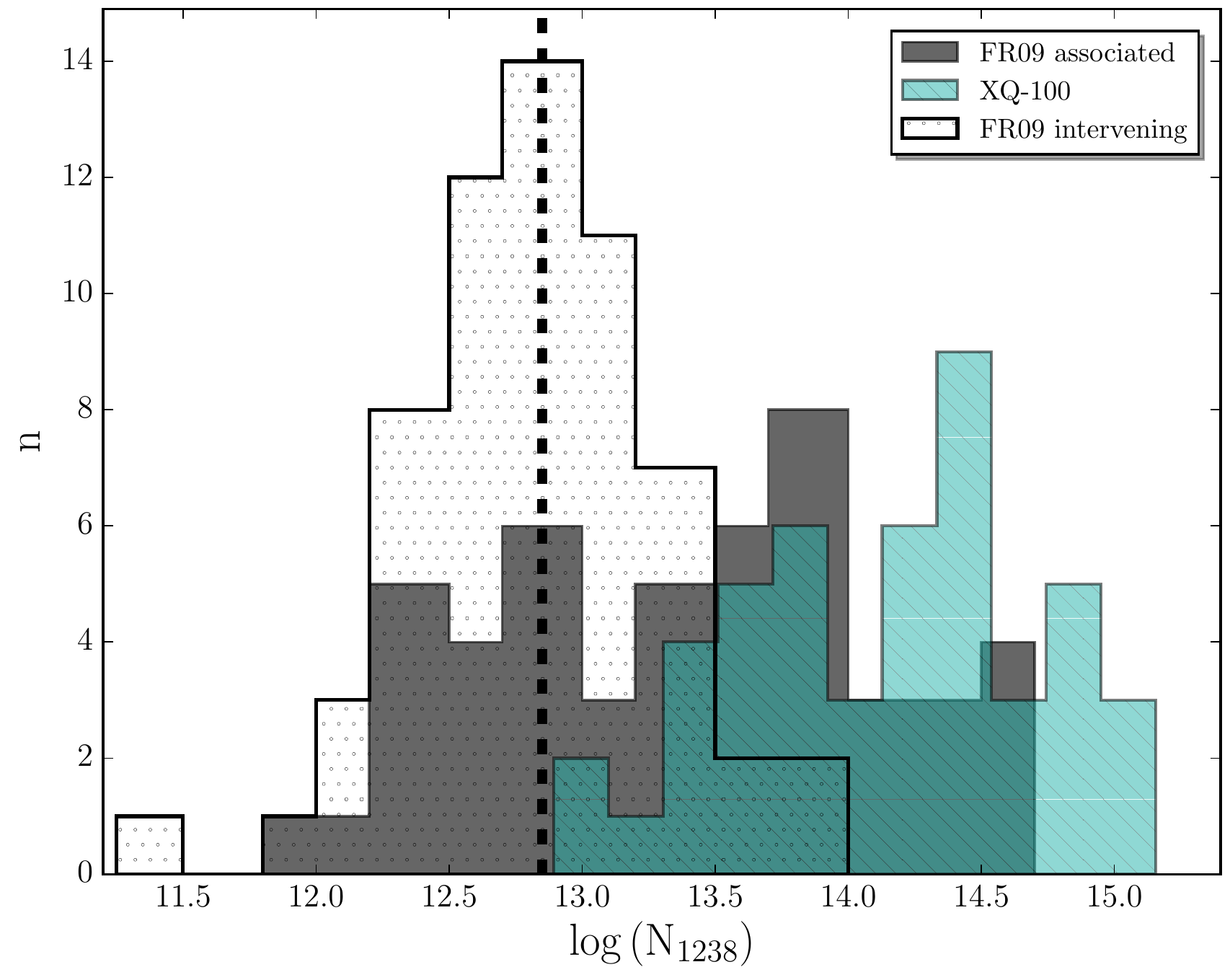}
  \caption{Distribution of the measured NV column densities of our sample (diagonal filled histogram). Also shown are the intervening (dotted histogram) and associated (dark shaded histogram) NV column density distributions from FR09. The dashed vertical line represents the detection limit of our sample, explaining the lack of low $\rm N_{1238}$ absorbers with respect to FR09 based on high resolution data.}
    \label{col_hysto}
\end{figure}

Motivated by these results, we explore in more detail the properties of our sample of NV absorbers.
Fig.~\ref{col_ratio} represents the velocity offset distribution of the NV and CIV column density ratios of our sample. 
 It is clear from Fig.~\ref{col_ratio} that most ($\sim 68$ percent) of the detections have a column densities ratio larger than 1. This is consistent with the result by FR09 : associated NV systems exhibit similar or even larger column density with respect to CIV. The fraction of quasars showing at least one NV absorption line with $\rm N_{1238}/N_{1548}>1$ is 26 percent.

Fig.~\ref{col_hysto} shows the distribution of the NV column densities of our sample (diagonal filled histogram), compared with the column density distributions by FR09. It is clear from Fig.~\ref{col_hysto} that  associated  NV systems start dominating the distribution at values of the order of $\rm log\, N_{1238}\sim13.5$ and that no intervening NV system with $\rm log\, N_{1238}>14$ is detected. Looking at our sample, 85 percent of the NV absorbers have column densities larger than  $\rm log\, N_{1238}=13.5$ and 55 percent have values larger than $\rm log\, N_{1238}=14$.  The number of quasars showing at least one NV absorption line with $\rm log\,N_{1238}>13.5$ is 30 percent, which reduces to 19 percent if a threshold of $\rm log\,N_{1238}>14$ is considered.

Furthermore, Kuraszkiewicz \& Green (2002) found that the NV/CIV ratio for the broad emission line (BEL) gas correlates strongly with the NV/CIV ratio for the associated NAL systems, while the control sample of intervening NV absorbers in their analysis does not show this correlation. Their finding identifies an additional test for the intrinsic nature of NALs in any given object. We will explore this correlation for our sample in a forthcoming paper (Perrotta et al. in prep.).

\subsection{Characterizing the quasar radiation field}

We have shown in the previous sections that we detect a statistically significant excess over the random occurrence of NALs up to $\rm 10,000\;km\;s^{-1}$ from the systemic redshift. This excess does not show any dependence on the quasar bolometric luminosity.

Looking at the CIV and SiIV velocity offset distributions relative to the quasar $\rm z_{em}$, we can appreciate the effect of the quasar radiation field influencing the gas  along the propagation direction of the outflows.
In particular, strong and weak lines in Fig.~\ref{offset_vel} (left bottom panel) display a different behaviour in the velocity range $\rm -1000 <v_{abs}< 5000\; km\,s^{-1}$. Indeed, the excess is less significant for the weaker lines. A possible explanation could be that in the first bin we examine gas closer to the emission redshift and therefore more directly influenced by the quasar radiation. 
Closer to the quasar, the gas is more easily ionized and it is thus reasonable to expect that the occurrence of the weaker lines  decreases there.
Furthermore, in general we find a less significant SiIV excess with respect to CIV. This is probably due to the different ionization potentials needed to produce these ions: $\rm 33.5$ and $\rm 47.9\;eV$, respectively. It is reasonable to expect that closer to the quasar, where the radiation field is more intense, a smaller fraction of Si is in the form of SiIV.
We have shown in the previous section that intrinsic NV systems exhibit similar or even larger column density with respect to CIV. The strong NV could be explained by an ionization effect but could also trace higher metallicities and N/C abundances characteristic of quasar outflows and environment (e.g. Wu et al. 2010). The equivalent width of NV lines could also be boosted relative to other transitions with larger column densities as an effect of partial coverage.
For what concerns the weaker lines, they are probably associated to galaxy halos in the quasar vicinity.

 We compared our results on the quasar proximity environment along the line of sight with the results obtained in the transverse direction using quasar pairs at close angular separation on the sky (e.g., Prochaska et al. 2014,  PR14 herafter).
Recent studies (e.g., Hennawi et al. 2006; Prochaska et al. 2013; Farina et al. 2013) have revealed that quasar host galaxies exhibit strong absorption features due to cool gas (e.g. $\rm HI, MgII, CII$) in the transverse direction. At impact parameters of $\rm \sim 100\; kpc$, the incidence of optically thick gas is  $> 50$ percent and the gas is substantially enriched (Prochaska et al. 2013; Farina et al. 2014). 
This is in contrast with what we observe along the line of sight, suggesting a scenario where the ionizing emission of the quasar is anisotropic (as predicted in AGN unification models). We can exploit these studies to evaluate if the excess we find is statistically significant over the environmental absorption, comparing our results with those derived from gas not directly influenced by quasar outflows. 

PR14 clearly demonstrate that the environment surrounding the host galaxies of luminous $\rm z\sim2$ quasars exhibit an excess of CIV absorption lines with $W_{1548}>0.2$ \AA\, to scales of $\rm 500 \; kpc$ from the quasar decreasing smoothly toward 1 Mpc. The authors quantify this excess by estimating the two-point-correlation function between CIV absorbers and quasars. The large clustering amplitude ($\rm r_0=7.5^{+2.8}_{-1.4}\,h^{-1}\;Mpc$) obtained from this analysis, implies that the CIV gas traces the same large-scale over densities as the halos manifesting $\rm z\sim2$ quasars. This result has been interpreted as evidence that the CIV gas is physically related to the massive halos ($\rm M>10^{12}\;M_{\odot}$) of galaxies that cluster with the quasar host.
Clearly, we cannot directly compare our detection fraction (Fig.~\ref{first_cov}, left) with the one derived in the PR14 analysis (see their Fig. 5, right). In fact, the detected CIV velocity offset  cannot  be directly translated into a physical distance because the redshift includes also the information on the peculiar velocity of the absorber. If we did that, we would obtain a proper separation of the order of $\rm \sim 10\; Mpc$, implying a lack of correlation between the absorber and the quasar host galaxy. Probably, all the gas studied by PR14 is contained in our first bin, within $\rm 5000\;km\;s^{-1}$ of $\rm z_{em}$. Indeed the average of their detection fraction (i.e. 0.51) is close to the value in our first bin (i.e. 0.43).
In conclusion, we observe a comparable incidence of CIV absorptions along and across the line of sight, suggesting that this ion will not allow us to disentangle AGN outflow systems from environmental absorptions.

A completely different behavior is shown by NV NALs. In section \S 4.4, we reported that, in our sample, NV exhibits an excess over random occurrence within $\rm 5000\;km\,s^{-1}$ of $\rm z_{em}$. Conversely, in the transverse direction only one NV detection has been detected in a sample of about 400 quasar pairs (Lau et al. 2015). This is further evidence that NV is the byproduct of the quasar ionizing radiation acting along the line of sight. 
 We reported in sections $\S4.2$ and $\S4.4$ that the excess of CIV gas at low velocities lies within the host halo and that most of this excess can be explained by the subset of CIV absorptions with detected NV. We are not able to infer the physical separation of the absorbing gas from the quasar, but NV NALs likely originate at small distances from the central engine (Wu et al. 2010). This absorbing material might be part of outflows in the accretion disk wind itself and its density would be order of magnitude bigger than that of intervening systems. 
In a future work we will perform detailed photoionization models to infer the physical conditions of some detected NV NALs in our sample exploiting high-resolution spectra obtained with UVES/VLT.

In addition, in the transverse direction a huge amount of CII with $\rm W_{1334}>0.2$ \AA\, extending to $\rm 200\;kpc$ from the considered quasars was found by PR14, (see their Fig. 5, left). The steep drop in covering fraction, $\rm f_C^{1334}$, at distances larger than $\rm 200\;kpc$ requires that this CII gas lies predominantly within the host halo.
In contrast with this result, we have shown in section \S 4.4,  (Fig.~\ref{first_cov}, right), that the frequency with which CII is detected in our survey is very small. Moreover, if we consider only absorbers with $\rm W_{1334}>0.2$ \AA, then we see that they are all related to identified DLA (or sub-DLA) systems probably unrelated with the host halo. Hence, we verify the absence of cool gas (in particular CII) along the line of sight associated to the quasar host galaxy, in contrast with what is observed in the transverse direction.

Wild et al. (2008) used a cross-correlation analysis of quasar-absorber pairs to measure the strength of narrow absorber clustering around quasars. They claim that galaxies in the vicinity of the quasar may contribute as much as $\sim55$ percent of the excess of the absorbers with $\rm v_{abs}\lesssim3000\,km\,s^{-1}$. We will perform our own CIV clustering analysis to validate the observed excess in a forthcoming paper (Perrotta et al. in prep.).

\section{Conclusions}
We have exploited the spectra of 100 quasars at emission redshift $\rm z_{em} = 3.5 - 4.5$ to construct a large, relatively unbiased, sample of NAL systems and statistically study their physical properties. The observations have been carried out with VLT/X-shooter in the context of the XQ-100 Legacy Survey. The combination of high S/N, large wavelength coverage and intermediate resolution makes XQ-100 a unique dataset to study NALs of high-z quasars in a single, homogeneous and statistically significant sample.
Inspecting the whole dataset we have identified almost one thousand CIV systems, covering the redshift range $ 2.55 < z < 4.73$. Furthermore, the quality and wavelength extent of our spectra allow us to look for other common ions (NV, SiIV and CII) at the same redshifts of the detected CIV absorbers. Contamination by the Ly-$\alpha$ forest prevent us from exploring the same velocity range for all the ions. In particular, these data allow us to look for NV only within $\rm 5500\; km \,s^{-1}$ of $\rm z_{em}$. The final sample contains 986 CIV, 236 SiIV, 46 NV and 28 $\rm CII_{1334}$ NAL detections.
Our primary results are as follows.

\begin{itemize}
\item[1)] The CIV sample exhibits a statistically significant excess ($\rm \sim 8\,\sigma$) within $\rm 10,000\; km \,s^{-1}$ of $\rm z_{em}$ with respect to the random occurrence of NALs, which is particularly evident for lines with $\rm W_0<0.2$ \AA. The excess is expected, but extends to larger velocities than the standard $\rm 5000\; km \,s^{-1}$ velocity cut-off usually adopted to identify associated system. The observed excess in the offset velocity distribution is not an effect of the NAL redshift evolution, but is likely due to quasar environment.
This excess is detected with a larger significance with respect to previous results and it does not show a significant dependence from the quasar bolometric luminosity. 
 Therefore, we suggest to modify the traditional definition of associated systems when also weak absorbers  ($\rm W_0<0.2$ \AA) \,are considered, extending to $\rm 10,000\; km \,s^{-1}$ the velocity cut-off mostly adopted in the literature.

\item[2)]  The CIV covering fraction, defined as the ratio between the number of quasars exhibiting at least one absorber with $\rm W_0 > 0.2$ \AA\, and the total number of quasars within a given velocity offset of $\rm z_{em}$, has the same value for $\rm v_{abs} < 5000\; km \,s^{-1}$ and $\rm v_{abs} < 10,000\; km \,s^{-1}$ of $\rm z_{em}$. Furthermore, we confirm that each quasar showing an absorber with $\rm 5000\; km \,s^{-1}< v_{abs}  < 10,000\; km \,s^{-1}$ has always at least an absorber with $\rm v_{abs} < 5000\; km \,s^{-1}$, while the opposite is not always true. This indicates the presence of a complex velocity structure, increasing the probability that these absorbers are part of outflows.

\item[3)] Out of the ions studied in this work, NV is the ion that best traces the effects of the quasar ionization field, offering an excellent statistical tool for identifying intrinsic systems. 
Based thereon, we compute the fraction of quasar hosting at least one intrinsic absorption system using the detection of NV corresponding to CIV absorptions in our sample to be 33 percent. This fraction is consistent with previous determinations based on other techniques (e.g., Misawa et al. 2007; Ganguly et al. 2013).

\item[4)]  Most of the NV lines in our sample have properties compatible with those of the intrinsic ones (see section $\S5$): 85 percent of the NV absorbers have column densities larger than $\rm log\,N_{1238}>13.5$ and 55 percent have values larger than $\rm log\,N_{1238}>14$.  In addition, 68 percent of the NV systems have column densities larger than the associated CIV. If we consider the subsample of NV with $\rm N_{1238}/N_{1548}>1$ the fraction of quasars with at least one intrinsic absorption-line becomes 26 percent. Alternatively, if we consider only systems with $\rm log\,N_{1238}>13.5$ and $\rm log\,N_{1238}>14$ we get values of the order of 30 and 19 percent, respectively.

\item[5)] Considering the radio properties of our sample the results of this work are in general agreement with those by Vestergaard (2003), although the latter sample has approximately an equal number of RLQs and RQQs, while our sample is almost entirely composed of RQQs. Taking into account the Vestergaard (2003) findings, we predict the number of CIV NAL with $\rm W_{1548} > 1.5$ \AA\,to be 2. We do find exactly 2 absorbers. 
The number of quasars showing CIV NALs in this work ($\rm \sim 72$ percent) is comparable
to the frequency ($\rm >60$ percent) observed by Vestergaard (2003). Finally, considering only NALs with $\rm W_{1548} > 0.5$ \AA\,we find exactly the same percentage (25 percent) of quasars with at least one associated CIV absorber within $\rm v_{abs} < 5000\; km \,s^{-1}$ of $\rm z_{em}$ as does Vestergaard (2003).

\item[6)] We verify the absence of cool gas (in particular CII) and the presence of highly ionized gas (traced by NV) along the line of sight associated to the quasar host galaxy, in contrast with what is observed in the transverse direction (see PR14 and section $\S 5.1$ of this current work). 
This result suggest a scenario where the ionizing emission of the quasar is anisotropic (as predicted by AGN unification models). 
\end{itemize}

The results of this work show strong evidence that NV is the byproduct of the quasar ionizing radiation acting along the line of sight. Therefore, it seems reasonable to utilize the presence of NV related to CIV absorbers to select the best outflow/intrinsic candidate NALs. These results, therefore, inspire the following lines of future inquiry:
\begin{enumerate}
\item A study to confirm that with the adopted selection criteria we are detecting all the NV intrinsic systems. With this aim, a series of photoionization models (with the code Cloudy, Ferland et al. 2013) will be performed of all the systems showing NV in our sample (Perrotta et al. in prep.).

\item A partial coverage analysis of the lines detected in our sample to be compared with the results of this current work. We will also perform a CIV clustering analysis to validate the observed excess outlined in item (1) above (Perrotta et al. in prep.).

\end{enumerate}

\section*{Acknowledgements}
We thank an anonymous referee for stimulating and detailed comments. The XQ-100 team is indebted to the ESO staff for support through the Large Program execution process, and to the DARK cosmology centre for financial support of the XQ-100 team meeting. SP, VD, GC and IP acknowledge support from the PRIN INAF "The X-Shooter sample of 100 quasar spectra at $\rm z\sim3.5$: Digging into cosmology and galaxy evolution with quasar absorption lines". JXP is supported by NSF grant AST-1109447. SLE acknowledges the receipt of an NSERC Discovery Grant. The Dark Cosmology Centre is funded by the DNRF. SL has been supported by FONDECYT grant number 1140838 and partially by PFB-06 CATA. K.D.D. is supported by an NSF AAPF fellowship awarded under NSF Grant AST-1302093. MV gratefully acknowledge support from the Danish Council for Independent Research via grant no. DFF 4002-00275.  SP thanks UCSC for the financial support and for hosting her during the development of part of this work. SP also thanks Lucia Armillotta, W. G. Mathews and Mauro Valli for the useful discussions.




\bibliographystyle{mnras}
\nocite{*}
\bibliography{biblio} 





\appendix

\section{Some identified absorbers}

  \begin{figure*}
  \centering
  \includegraphics[width=8cm]{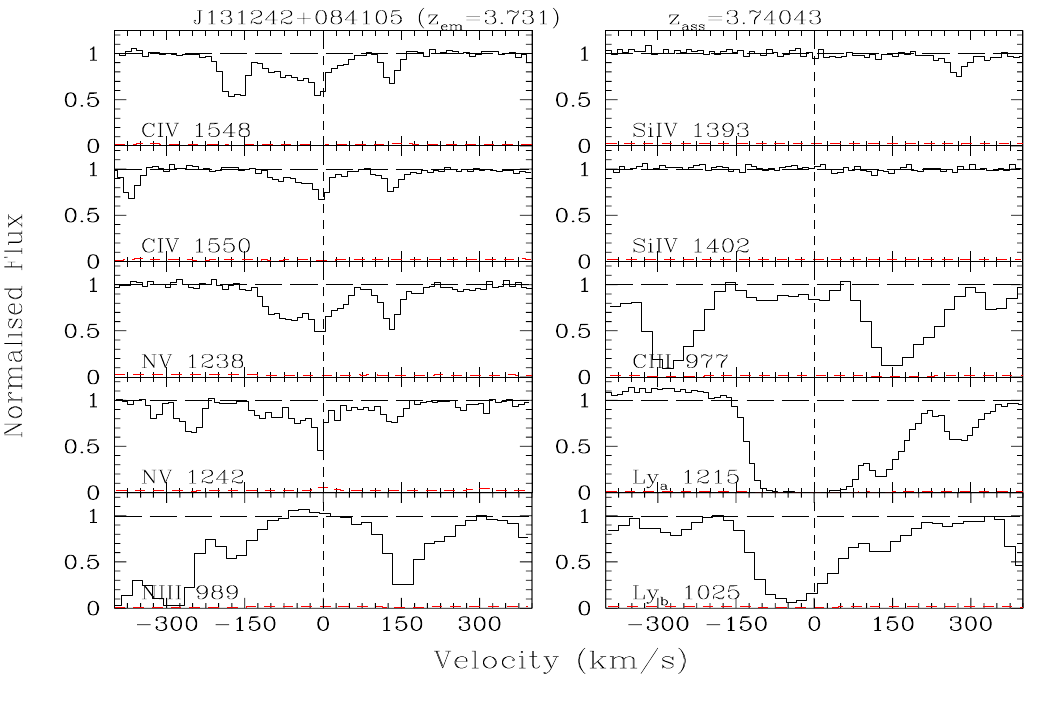}
  \qquad\qquad
    \includegraphics[width=8cm]{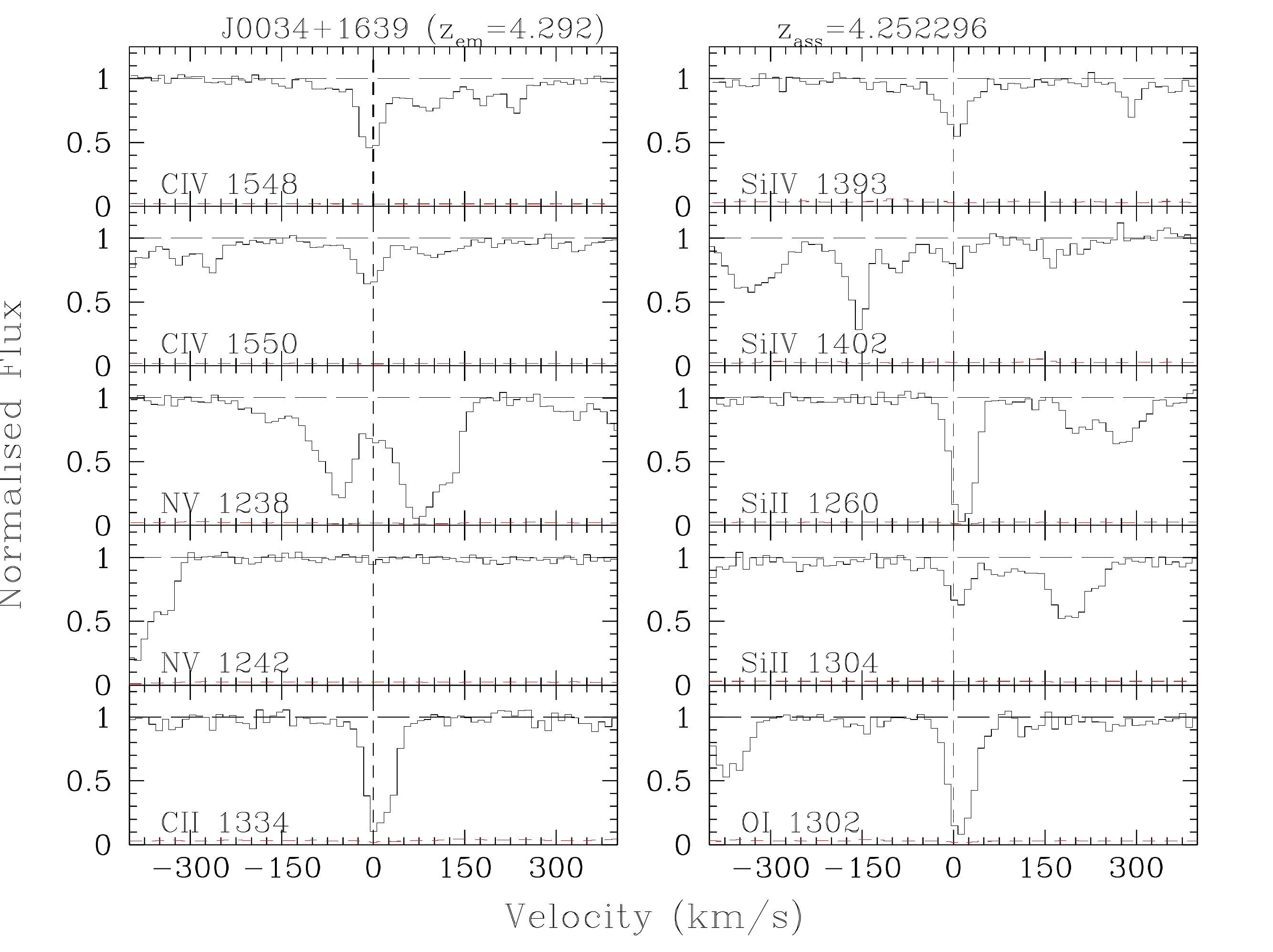}
  \includegraphics[width=8cm]{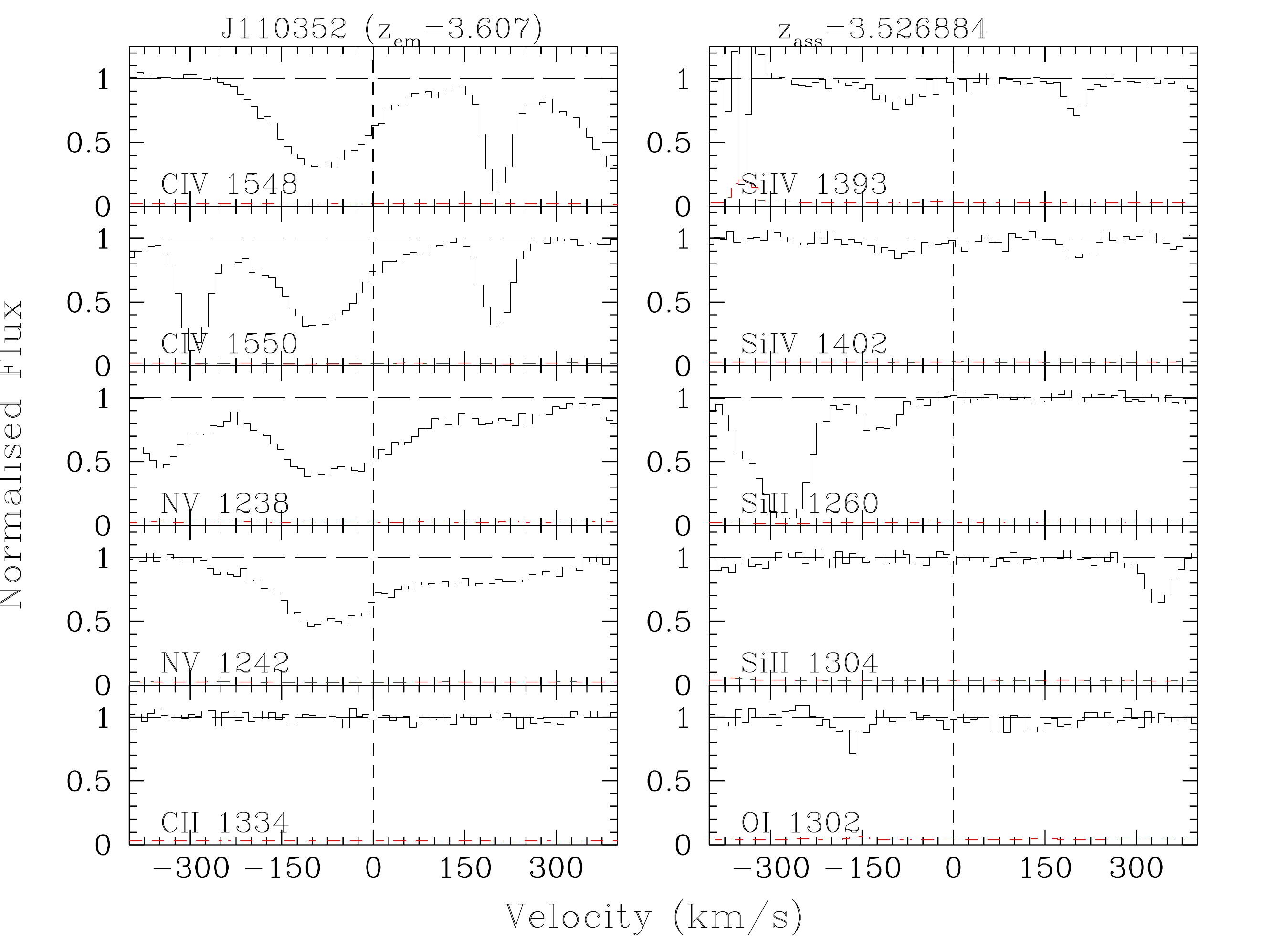}
  \qquad\qquad
  \includegraphics[width=8cm]{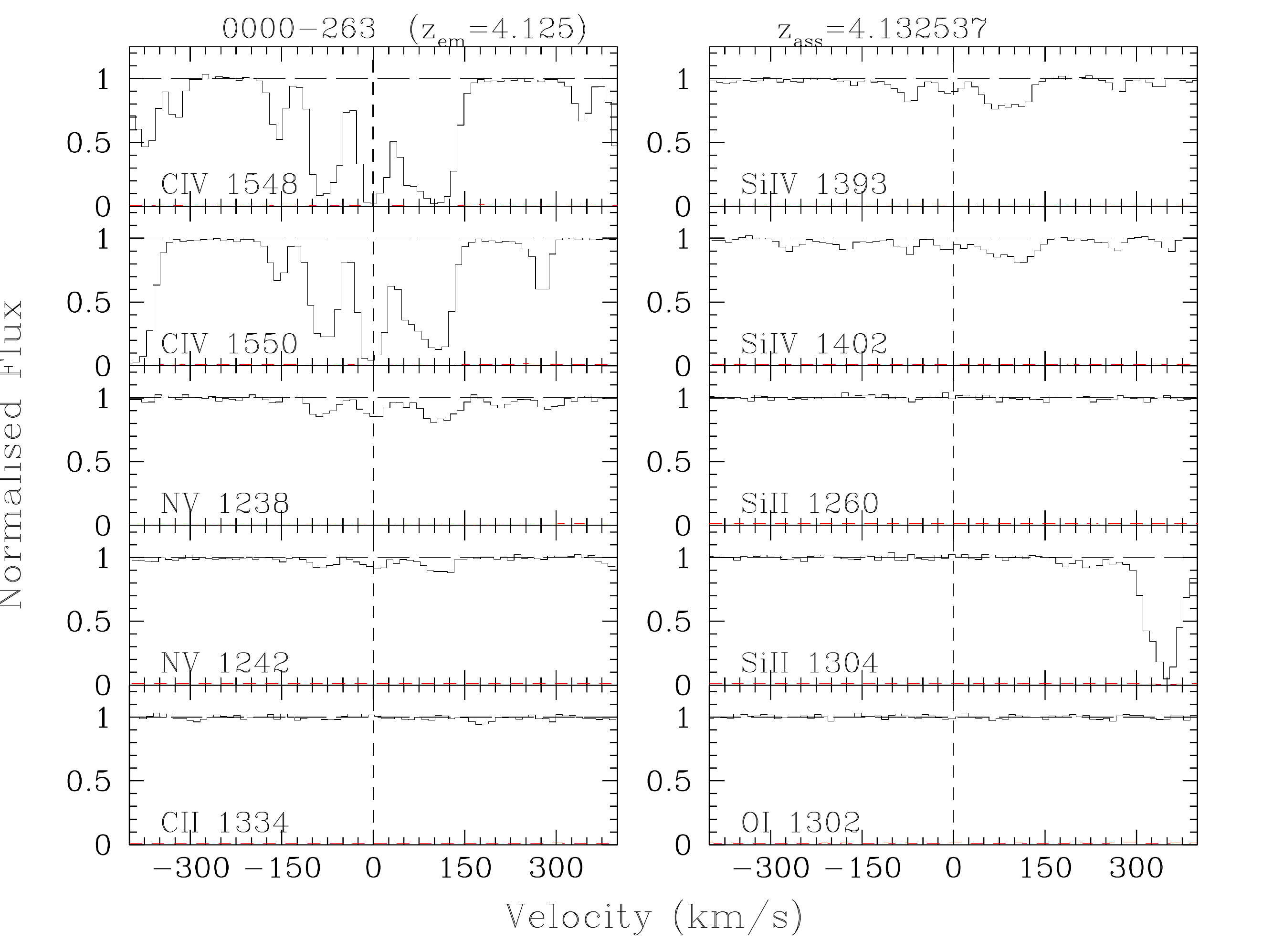}
  \caption{Some examples of the identified absorbers.} 
  \label{J1103}
\end{figure*}


\bsp	
\label{lastpage}
\end{document}